\documentstyle[preprint,eqsecnum,prb,aps,amsfonts,amssymb,epsfig]{revtex}

\newcommand{\bleq}{\ifpreprintsty
                   \else
                   \end{multicols}\widetext \vspace*{-3.5ex}{\tiny
                   
                \noindent\begin{tabular}[t]{c|}
                   \parbox{0.493\hsize}{~} \\ \hline \end{tabular}}
                                      \fi}
\newcommand{\eleq}{\ifpreprintsty
                   \else
                   {\tiny\hspace*{\fill}\begin{tabular}[t]{|c}\hline
                    \parbox{0.49\hsize}{~} \\
                    \end{tabular}}\vspace*{-2.5ex}\begin{multicols}{2}
                    \narrowtext
                    \fi}
\newcommand{\bcols}{\ifpreprintsty\else\begin{multicols}{2} 
        \narrowtext\fi}
\newcommand{\ecols}{\ifpreprintsty\else\end{multicols}\fi}

\begin{document}


\draft
\tighten
\title{Lattice models and Landau theory for type II incommensurate
crystals}
\author{G. H. F. van Raaij,\cite{guido} K. J. H. van Bemmel,\cite{koert} 
and T. Janssen}
\address{Institute for Theoretical Physics,
University of Nijmegen, Toernooiveld 1, NL-6525 ED Nijmegen, The Netherlands}
\maketitle

\begin{abstract}
Ground state properties and phonon dispersion curves of a classical linear 
chain model describing a
crystal with an incommensurate phase are studied. This model is 
the DIFFOUR (discrete frustrated $\phi^4$) model with
an extra fourth-order term added to it. The incommensurability in these
models may arise if there is frustration between nearest-neighbor and
next-nearest-neighbor interactions. We discuss the effect of the
additional term on the phonon branches and phase diagram of the DIFFOUR model.  We find some features not present in the DIFFOUR model such as the 
renormalization of the nearest-neighbor coupling. Furthermore the ratio
between the slopes of the soft phonon mode in the ferroelectric and
paraelectric phase can take on values different from $-2$. Temperature 
dependences of the
parameters in the model are different above and below the paraelectric
transition, in contrast with the assumptions made in Landau theory.
In the continuum limit
this model reduces to the Landau free energy expansion for type II
incommensurate crystals and it can be seen as the lowest-order generalization
of the simplest Lifshitz-point model. Part of the numerical calculations 
have been done
by an adaption of the Effective Potential Method, originally used for
models with nearest-neighbor interaction, to models with also 
next-nearest-neighbor interactions.
\end{abstract}




\section{Introduction}

Like other transitions, the phase transition to an incommensurate (INC)
phase (reviews are given by Bak, \cite{bak} Selke \cite{selke} and Janssen 
and Janner \cite{tedalo}) can be described on the phenomenological level 
within the frame of extended Landau theory. \cite{toledano}
The necessary extension consists essentially in
accounting for the expansion of the free energy density as a function not
only of the components of the order parameter, but also of their spatial
derivatives. Therefore the global free energy becomes a functional of
spatially dependent components of the order parameter and the equilibrium
configuration for given values of temperature and external parameters is
found as a solution of a variational problem.

The continuum Landau theory allows a natural classification 
\cite{brucecowleymurray} of the possible
forms of the free-energy functional for an INC transition into two classes,
according to whether the driving term in the free energy expansion
responsible for the appearance of the incommensurate state is linear (type
I, Lifshitz invariant present) or quadratic (type II, no Lifshitz
invariant) in the gradient of the order parameter. The properties of
those two kinds of INC phases are different: for type I INC phases the
lock-in transition is either continuous, or only slightly discontinuous,
and approaching the lock-in temperature $T_{c}$ these phases exhibit the 
structuration of the modulated phase into discommensurations or solitons. 
On the other hand, the modulation of the type II INC phase remains 
practically sinusoidal down to the temperature $T_{c}$ and the lock-in 
transition is always of first order. Although the above statements can be 
considered as a rule of thumb, there are cases known where there is
coexistence of solitonic and sinusoidal structural modulation. See Aramburu
\cite{aramburu} for details.

In the following we will only be concerned with models describing type II
INC phases. Landau theory has been rather successful in describing basic 
properties of these phases, but if one wants to have a better 
understanding of the true microscopic origin of the INC phase one has to go
beyond this phenomenological approach. One possibility would be to study
full microscopic models with realistic interactions. Another approach,
to which the main part of this paper will be devoted, is
to study semi-microscopic models which take into account the discrete
nature of the systems and discuss properties in terms of (effective)
inter-atomic interactions. 

The discreteness of a lattice leads to a number of important physical 
consequences such as  pinning of solitons in the anisotropic 
next-nearest-neighbor Ising (ANNNI) model 
(see Yeomans \cite{yeomans} for a review) and in the Frenkel-Kontorova model, 
\cite{pokrovsky} and the existence of a devil's staircase (infinite 
number of commensurate and incommensurate phases), for example found in 
betaine calcium chloride dihydrate (BCCD). \cite{brillplusunruh} Within
Landau theory it is difficult to explain the occurrence of a specific
sequence of transitions: several lock-in terms are needed then.
For example,
Ribeiro {\it et al.} \cite{ribeiro} chose the magnitude of four distinct
lock-in contributions to the free energy in such a way as to stabilize the
four most prominent commensurate phases in BCCD.
Furthermore, in discrete models chaotic states are possible, 
\cite{bak,wreszinski} which may provide an alternative description of
phenomena observed in for example spin glasses, superionic conductors, 
the magnetic system CeSb and systems with pinning of charge density waves. 
See Bak \cite{bak} for details.

In the past
few years different lattice models have been constructed, for example to
describe the phase transitions in the A$_2$BX$_4$ family,
\cite{chenwalker} in BCCD \cite{chenwalkerplusfolkins} and,
more general, in
crystals with $Pcmn$ symmetry. \cite{ted1} These models are 2-dimensional,
with only nearest-neighbor interactions. Hlinka {\it et al.} \cite{jirka}
studied a 3-dimensional nearest-neighbor model, applicable to BCCD. All
these models have in common that the frustrated interaction, needed for
having an incommensurate phase, comes from a nearest-neighbor mixing
interaction. 

Recent X-ray, \cite{israel} neutron \cite{stephans6,stephanse6} and Raman
\cite{stephanraman} experiments on the Sn$_2$P$_2$(S$_{1-x}$Se$_x$)$_6$ 
crystal family of uniaxial ferroelectrics
motivated us to study lattice models. In the
composition-temperature phase diagram of this crystal family a Lifshitz
point is present.\cite{vysochanskii} At this point the paraelectric phase, 
the ferroelectric phase and the INC phase become equal, and the boundaries
separating these phases have equal derivative.
Such a special point was, except for some
ferroelectric liquid crystals, only found in the temperature-applied
magnetic field phase diagram of the magnetic compound MnP. \cite{becerra}
From both experimental and theoretical point of view this uniaxial Lifshitz     point is
interesting because critical exponents deviate \cite{folk} from those
found for ordinary critical points. This crystal family furthermore
displays an interesting modulation wave vector behavior, 
shows cross-over effects from order-disorder to a displacive type of phase 
transition, \cite{jirka2} and the ratio between the slopes of the soft
phonon mode in the ferroelectric and paraelectric phase deviates
\cite{stephans6} substantially from the standard value $R= -2$. These 
features are much easier to understand in a lattice model than in Landau 
theory.

The paper is arranged as follows:
in Section II we present a one-dimensional model; in many anisotropic 
systems, like Sn$_2$P$_2$(S$_{1-x}$Se$_x$)$_6$, the modulation wave vector 
is in one
specific direction. The incommensurability may arise if there is frustrating
interaction between nearest-neighbor and next-nearest-neighbor couplings.
We discuss general features of the model. In Section III we give some exact     results regarding ground state properties.
In Section IV we discuss the dynamics and the stability of the
various phases.  In Section V the phase diagrams, calculated partly
analytically, partly numerically, are presented. Temperature effects are 
treated in Section VI. In Section VII we discuss the continuum limit of the 
model which gives the connection with the Landau theory. We conclude and
give an outlook for further research in Section VIII.
In Appendix A we give some exact results for phase boundaries
and in Appendix B we present the next-nearest-neighbor extension of 
the Effective Potential Method for the determination of the ground state. 
This method was used to calculate some of the phase diagrams.

\section{An extension of the DIFFOUR model}

In the following we will be concerned with an extension of the so-called
DIFFOUR model \cite{ted2} (discrete frustrated $\phi^4$ model), for which
the potential energy can be written as
\begin{eqnarray}
V =  \sum_n \Big\{&&\frac{A}{2}x_n^2 + \frac{B}{4}x_n^4 
+ \frac{C}{2}\left( x_n - x_{n-1} \right)^2 + \frac{D}{2} \left( x_n - 
x_{n-2} \right)^2 \nonumber \\
&&+\frac{E}{2} \left[ x_n^2 \left(x_n - x_{n-1}\right)^2 + x_n^2
\left(x_n - x_{n+1} \right)^2 \right] \Big\}.
\label{edif}
\end{eqnarray}
The original DIFFOUR model, or EHM (elastically hinged molecule) model,
\cite{dmitriev} has $E=0$. Although in principle this model
gives incommensurate ground states, the behavior of the modulation wave
vector as found in experiments can not be reproduced satisfactorily by the
model. In order to account for this shortcoming we supply the DIFFOUR model
with a non-linear coupling to neighbors. There are several possibilities:
if we restrict ourselves to fourth-order terms we can consider a term
$\propto (x_n - x_{n-1})^4$. The resulting model has been studied by Lamb,
\cite{lamb} who showed that the origin of this term is related to strain
terms in the so-called magnetoelastic DIFFOUR model.
Another possibility would be to consider a term of the form $\propto (x_n -
x_{n-2})^4$. Without the term mentioned above this would however be rather
unphysical. Instead we will choose a term of the form $\propto x_n^2(x_n -
x_{n \pm 1})^2$. This is the lowest-order dispersive fourth-order term, as 
will be shown in Section VII, and can, for example, be obtained from strain     terms.

The order parameter $x_n$ can be, for example, a displacement, a component 
of the polarization $P$ for ferroelectric systems, a component of the 
magnetization $M$ for magnetic systems, a rotation angle or a strain 
component. In this article we use for convenience terms like paraelectric,
ferroelectric and antiferroelectric to distinguish between different ground
states. The origin of incommensurability in this model is essentially
competition between interactions with nearest- and next-nearest neighbors  
which may lead to frustration. Higher order terms are needed for 
stabilization.

We expect that 
the extra fourth-order term $(E \neq 0)$ has a large effect on the phase
diagrams for $E=0$. To actually determine phase diagrams it is not 
necessary to vary all 5 parameters $A,B,C,D,E$, which can be seen as 
follows: by taking $x_n' = \sqrt{B/|D|} x_n$ and $V' = B/|D|^2 V$
we get the following renormalized parameters: $A' = A/|D|$, $B' = 1$, $C' =
C/|D|$, $D' = D/|D| = \pm 1$, and $E' = E/B$.

For some purposes it is convenient to rewrite the potential in the
following form:
\begin{eqnarray}
V = \sum_n \Big\{&& \frac{\tilde{A}}{2}x_n^2 + \frac{\tilde{B}}{4}x_n^4 +
\tilde{C} x_n x_{n-1} + \tilde{D} x_n x_{n-2} \nonumber \\
&&+ \frac{\tilde{E}}{2} 
\left[ x_n^2 \left(x_n - x_{n-1}\right)^2 + x_n^2 \left(x_n - x_{n+1} 
\right)^2 \right] \Big\},
\label{deftilde}
\end{eqnarray}
with $\tilde{A} = A + 2C + 2D$, $\tilde{B} = B$, $\tilde{C} = -C$,
$\tilde{D} = -D$ and $\tilde{E} = E$. From this the connection with the
ANNNI model can easily be made. Let us put $\tilde{E} = 0$ and 
$\tilde{B} = - \tilde{A}$. If we now take the limit $\tilde{A} 
\rightarrow - \infty$ we end up with a model with two infinitely deep wells.
The $x_n$ can only take on values $\pm 1$ and can thus be seen as spins. 
These spins are coupled to nearest-neighbors and next-nearest-neighbors via
the $\tilde{C}$ and $\tilde{D}$ terms. So by increasing the depth of the
double-well potential there is a crossover from displacive behavior to
order-disorder behavior in the transition from the normal to the
incommensurate phase.

Inserting $x'_n = (-1)^n x_n$ in the above potential leads to
\begin{eqnarray}
V = \sum_n \Big\{&& \frac{\tilde{A}}{2}{x_n'}^2 + 
\frac{\tilde{B}}{4}{x_n'}^4 - \tilde{C} x_n' x_{n-1}' + \tilde{D} 
x_n' x_{n-2}' \nonumber \\
&&+ \frac{\tilde{E}}{2} \left[ {x_n'}^2 \left(x_n' + 
x_{n-1}' \right)^2 + {x_n'}^2 \left(x_n' + x_{n+1}'
\right)^2 \right] \Big\}.
\end{eqnarray}
In the DIFFOUR model ($\tilde{E} = 0$) this leads to the following 
symmetry: if $\{ x_n \}$ is a state for $\tilde{C} = X$, then 
$\{(-1)^n x_n \}$ is a state for $\tilde{C} = -X$ with the same energy. 
This property can for example be seen in the 
ferroelectric-antiferroelectric phases. However, for $\tilde{E} \neq 0$ this
symmetry $\tilde{C} \leftrightarrow -\tilde{C}$ is no longer present.

\section{Ground state properties}

Different ground states are possible, depending on the values of the
parameters. The stationary states are solutions of $\partial V / \partial
x_n = 0$, giving
\begin{eqnarray}
&& Ax_n + Bx_n^3 + C \left( 2x_n -
x_{n-1} - x_{n+1} \right) 
+ D \left( 2x_n - x_{n-2} - x_{n+2} \right) \nonumber \\
&&+ E \left[ 4x_n^3 - 3x_n^2 \left( x_{n-1} + x_{n+1} \right) 
 + 2x_n \left( x_{n-1}^2 + x_{n+1}^2 \right) -  \left( x_{n-1}^3 + 
x_{n+1}^3 \right) \right] = 0.
\label{statcon}
\end{eqnarray}
If we impose periodic boundary conditions $x_{N+n} = x_n$ we arrive at a
set of $N$ coupled non-linear equations. To find the lowest-energy state
for each solution and for each value of $N$ the potential energy has to be 
evaluated.
For low-period commensurate states (small values of $N$) analytic 
solutions of the above equation can be found. 
In the following, we study (for fixed $N$) periodic solutions of 
(\ref{statcon}).
For them we give the equilibrium values $\{x_n\}$ and the corresponding
energy per particle $v = V / N$. 

In the paraelectric state $(N=1)$ all particles are in the equilibrium 
positions
\begin{equation}
x_n = 0, \quad v = 0.
\label{defpar}
\end{equation}

In the ferroelectric state $(N= 1)$ all particles are uniformly
displaced from their equilibrium positions
\begin{equation}
x_n = \sqrt{ - \frac{A}{B} }, \quad v = - \frac{ A^2}{4 B}.
\label{deffer}
\end{equation}
Note that $B$ always has to be positive, for the potential to be bounded
from below. This implies that the ferroelectric state only can exist for
$A<0$. For $A>0$ the ground state may be paraelectric.

In the following we give some analytic results, based on numerical
calculations of the shape of the solution $\{x_n\}$.

In the antiferroelectric state $(N=2)$ particle positions alternate    
along the chain
\begin{equation}
x_n = ( - 1)^n \sqrt{ -  \frac{A + 4 C}{B + 16 E} }, 
\quad v = - \frac{ \left( A + 4 C \right)^2}{4 ( B + 16 E)}.
\label{defantifer}
\end{equation}
The potential is unbounded from below for $E \leq -B/16$ and stable
solutions exist only for $E > -B/16$ and $A + 4C < 0$. Both conditions have
to be satisfied.

For $N=3$ we determined the solution with lowest energy to be of    
the form $(x_1,x_2,x_3) = (k\xi,\xi,k\xi)$, with $x_1 / x_2 < 0$, and
\begin{equation}
\xi^2 =  - \frac{A + 2(C+D)(1-k)}{B + 2E(2 -3k + 2k^2 -k^3)} 
= - \frac{Ak + (C+D)(k-1)}{Bk^3 + E(2k^3 -3k^2 +2k -1)}.
\label{defxi}
\end{equation}
The factor $k$ is determined by
\begin{eqnarray} \label{k4}
&&2 \left[ B(C+D) + E(-A+C+D) \right] k^4 - \left[ B(A+2C+2D) + 2E(-A+2C+2D)
\right] k^3   \\
&& -3AE k^2 + \left[ B(A+C+D) + 2E(A+2C+2D) \right] k - \left[ B(C+D) + E(-A+2C+2D)
\right] = 0. \nonumber
\end{eqnarray}
The energy per particle is given in terms of $\xi$ and $k$ as 
\begin{equation}
v = \frac{\xi^2}{6} \left[ A(1+2k^2) + 2(C+D)(1-k)^2 \right] 
+ \frac{\xi^4}{12} \left[ B(1+2k^4) + 4E(1+k^2)(1-k)^2 \right].
\end{equation}
Note that the quartic equation (\ref{k4}) can be written as
\begin{eqnarray}
(1-k) \Big\{&& -2 \left[ B(C+D) + E(-A+C+D) \right] k^3 + \left[ BA + 2E(C+D)   \right] k^2 \nonumber \\
&& + \left[ BA + E(3A + 2C+2D) \right] k  - \left[ B(C+D) + E(-A +2C+2D) 
\right] \Big\} = 0, 
\end{eqnarray}
where the special solution $k=1$ gives a ferroelectric state.
The solution of the remaining cubic
equation, which can be solved exactly for given parameters, gives a $k$ such 
that $x_1 / x_2 < 0$, a true $N=3$ state.

The lowest energy state for $N=4$ has $x_1 = x_2 = \rho,~x_3 = x_4 = -\rho$     with 
\begin{equation}
\rho = \sqrt{- \frac{A + 2 C + 4 D}{B + 8E}}, \qquad v = - \frac{(A + 
2 C + 4 D)^2}{4 (B + 8 E)}.
\label{defrho}
\end{equation}
We have to keep in mind that we must satisfy $E > -B/16$, which is not 
obvious from the above expression, but comes from the analysis of the
antiferroelectric state.

The lowest energy solution for $N=6$ can analytically be obtained, in the
same manner as for $N=3$. It has the form $(x_1,x_2,x_3,x_4,x_5,x_6) =
(k\xi,\xi,k\xi,-k\xi,-\xi,-k\xi)$. The lowest energy solution for $N=8$ reads
$(x_1,x_2,x_3,x_4,x_5,x_6,x_7,x_8) = (k\xi,\xi,\xi,k\xi,-k\xi,-\xi,-\xi,
-k\xi)$. For $N=5$ one needs 2 different values of $k$:
$(x_1,x_2,x_3,x_4,x_5) = (k\xi,k'\xi,\xi,k'\xi,k\xi)$ and for $N=7$ one needs
three different values: $k,k',k''$, and the above method no longer works
for $N=5$ and $N=7$. Therefore, to find
states with larger periods, or even incommensurate periods,
we rely on numerical calculations, for which
true incommensurate states of course never can be found. However, the idea 
is that such a state can always be arbitrary 
well approximated by a commensurate state with wavelength
\begin{equation}
\lambda = \frac{N}{s}, \quad s = 1,2,\ldots, \quad N \geq 2s,
\end{equation}
where $N,s$ are coprime numbers. Such a solution has a period $N$ and, in
general, $2s =$ (number of local minima $+$ number of local maxima). In
the special case where the $\{x_n\}$ take on positive and negative values,
$2s =$ (number of sign changes within the period $N$).
The bigger $N$ and $s$, the better the approximation. 

\begin{figure}
\begin{center}
\epsfig{file=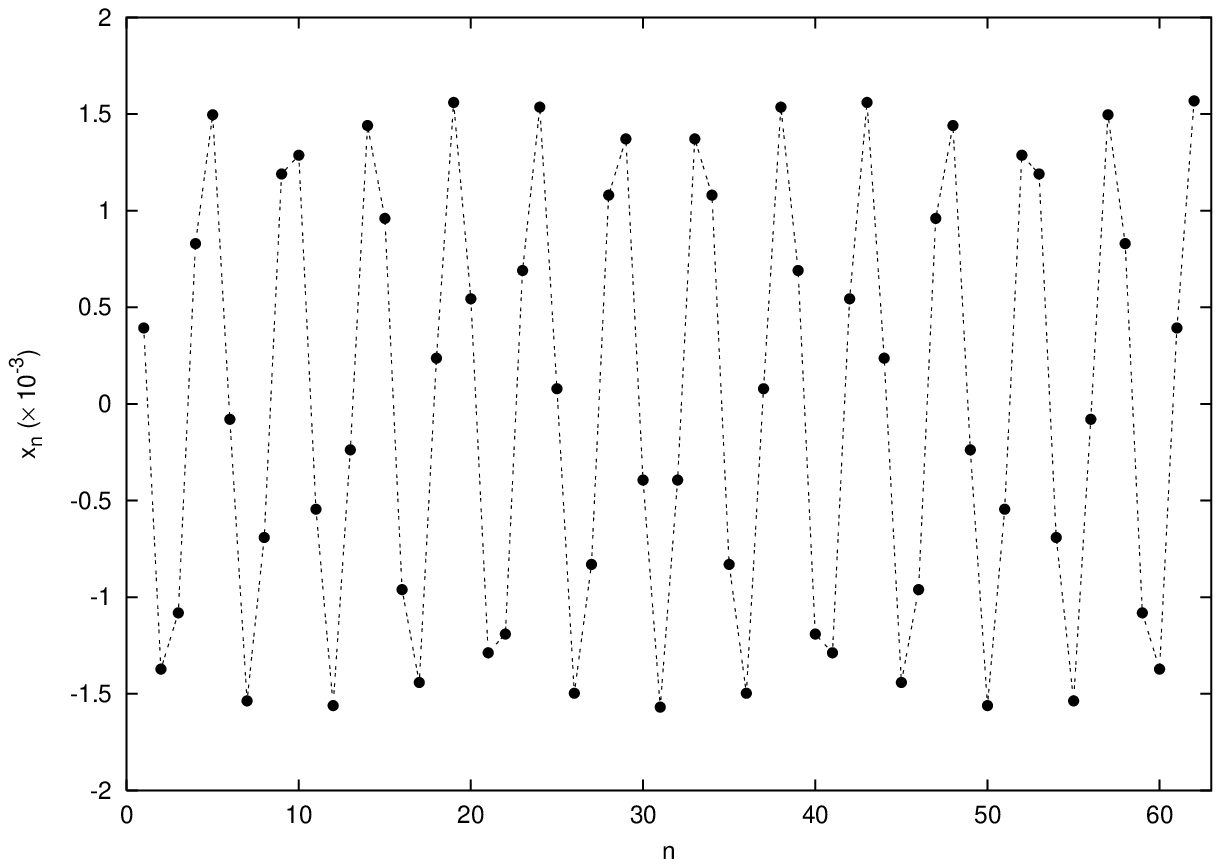,width=\linewidth}
\end{center}
\caption{Ground state configuration for $A=2.24999,~B=1,~C=1,~D=-1$
and $E=1$ in the commensurate approximation 62/13. $x_n$ is the
displacement for particle $n$. For $A,B,C,D,E$ given above
the system is just below the phase boundary between the
paraelectric phase and the incommensurate phase. Although the displacements
are small, ${\cal O}(10^{-3})$, the onset of the incommensurate phase
is evident. Points are calculations, lines are drawn to guide the eye.}
\label{grdstate62-13}
\end{figure}

As an example of a numerical calculation we consider the ground state for
$A=2.24999,~B=1,~C=1,~D=-1,~E=1$. It is known to be incommensurate (see 
Sections IV and V) with a wavelength $\arccos(\frac{1}{4}) \approx 
4.76679213$. By the 
Farey construction \cite{tedalo} we find that $\frac{62}{13} \approx 
4.76923077$ should be a reasonable
commensurate approximation. We numerically determined the ground state in
terms of the $\{x_n\}$. The result is shown in Figure \ref{grdstate62-13}.
After 62 particles the sequence repeats itself, and the solution passes 
through zero 26 times. We can label this state by its modulation wave
vector $\frac{13}{62}$, measured in units of $2 \pi$.

For certain regimes in the parameter space the ground state can be
determined analytically. The entire phase boundary of the paraelectric state    and a part of the phase boundary of the ferroelectric state can be
calculated. 
For the nearest-neighbor case in the DIFFOUR model proofs are given by
Janssen and Tjon. \cite{ted3} We extend their proofs to the case in which
we also have next-nearest-neighbor interaction. As the proof is rather
lengthy it will be given in Appendix A.

\section{Phonon dispersion curves and stability limits}

To decide whether a solution of the equilibrium conditions is locally stable
or not, one considers small displacements $\epsilon_n$ from the positions
given by a static solution $\{x_n\}$ satisfying (\ref{statcon}):
\begin{equation}
u_n = x_n + \epsilon_n.
\end{equation}
The phonon frequencies are given by the square roots of the eigenvalues
of the dynamical matrix and
for stability all eigenvalues have to be non-negative. The dynamical matrix
for a period-$N$ solution $(N\geq5)$ has elements
\begin{eqnarray}
D_{n,n} &=& A + 3B x_n^2 + 2C + 2D 
+ 2E \left[ 6 x_n^2 - 3x_n (x_{n-1} +
x_{n+1} ) + x_{n-1}^2 + x_{n+1}^2 \right], \nonumber \\
D_{n,n\pm1} &=& -C + E \left[ -3 x_n^2 + 4 x_n x_{n\pm1} - 3 x_{n\pm1}^2
\right],  \\
D_{n,n\pm2} &=& -D, \nonumber
\end{eqnarray}
with $x_{N+n} = x_n$ and the special cases
\begin{eqnarray}
D_{1,N} &=& \left[ -C + E \left( -3 x_1^2 + 4 x_1 x_N -3 x_N^2 \right)
\right] e^{-iq}, \nonumber \\
D_{1,N-1} &=& D_{2,N} = -D e^{-iq}.
\end{eqnarray}
In the above expressions the $\{x_n\}$ are solutions of (\ref{statcon}).
Furthermore $D_{n,m} = D_{m,n}^*$ and all other matrix elements are zero.

In the following the phonon branches for certain low-period states will be 
examined. This will be done in terms of $A,B,C,D,E$.

\subsection{Paraelectric state}

For the paraelectric state (\ref{defpar}) the 
dynamical matrix is given by
\begin{equation}
{\sf D} = A+2C(1-\cos (q))+2D(1-\cos (2q))=m\omega ^2.
\end{equation}
Rewriting this equation gives
\begin{equation}
m\omega ^2=A+(4C+16D)\sin ^2(\frac{q}{2})-16D\sin ^4(\frac{q}{2}).
\label{parphonon}
\end{equation}
Note that this expression does not contain $E$. This means that the
stability limits for the paraelectric phase in this model are the same as
those for the paraelectric state in the DIFFOUR model.
Now, we are looking for the minimum of this phonon branch.
We distinguish the cases $D>0$ and $D<0$.
The results are summarized in Table \ref{partab}.

For $D>0$ and $C =0$, the branch has two minima, at 
$\sin ^2(\frac{q}{2})=0$ and
$\sin ^2(\frac{q}{2})= 1$. For $A<0$ the ferroelectric state and the
antiferroelectric state are degenerate, for $E=0$ only.
Comparison with calculations in Appendix A shows that as long as the
paraelectric state is stable, it is the ground state. Destabilization is
the condensation of a soft phonon.

\subsection{Ferroelectric state}

For the ferroelectric state (\ref{deffer}) one has
\begin{eqnarray}
m\omega ^2&=&A+3B x^2+2C(1-\cos (q))+2D(1-\cos (2q)) 
+4E x^2(1-\cos (q)) \nonumber \\
&=&-2A+(4C+16D-\frac{8AE}{B})\sin ^2(\frac{q}{2}) 
-16D\sin ^4(\frac{q}{2}).
\label{ferphonon}
\end{eqnarray}
See Table \ref{fertab} for the analysis. Note that for $D>0$ there
is degeneracy for $E=0$.

\subsection{Antiferroelectric state}

Finally, the phonon branches of the antiferroelectric state 
(\ref{defantifer}) will be investigated.
The $2\times 2$ dynamical matrix is given by 
\begin{eqnarray}
D_{1,1} = D_{2,2} &=& A+2C+2D(1-\cos (q)) 
-(3B+28E)\frac{A+4C}{B+16E}, \nonumber \\
D_{1,2} = D_{2,1}^* &=& (e^{-iq}+1)(-C+10E\frac{A+4C}{B+16E}).
\end{eqnarray}
The eigenvalue equation reads
\begin{eqnarray}
m\omega ^2&=&A+2C+4D-(3B+28E)\frac{A+4C}{B+16E} 
- 4D\cos ^2(\frac{q}{2}) \nonumber \\
&&  \pm 2(-C+10E\frac{A+4C}{B+16E})\cos (\frac{q}{2}).
\label{antiferphonon}
\end{eqnarray}
Results are summarized in Table \ref{antifertab}. Note that for $C =
10E\frac{A+4C}{B+16E}$ the two branches coincide.

\subsection{States with period $N \geq 3$}

For the $N=3$ solution exact phonon frequencies can in principle be found. 
The elements of the dynamical matrix are given in terms of $\xi$ and $k$,
defined in (\ref{defxi}) and (\ref{k4}):
\begin{eqnarray}
D_{1,1} &=& D_{3,3} = A + 2(C+D) 
+ \left[ 3 B k^2 + 2E (4 k^2 -3k + 1) \right] \xi^2 , \nonumber \\
D_{2,2} &=& A + 2(C+D) + \left[ 3 B + 4 E (k^2 - 3 k + 3) \right] \xi^2 ,
\nonumber \\
D_{1,2} &=& D_{2,3} = -C -  E ( 3 k^2 -4k +3)\xi^2 -D e^{-iq}, \\
D_{1,3} &=& -D - ( C + 2 E k^2 \xi^2 ) e^{-iq}. \nonumber
\end{eqnarray}  
The eigenvalues are then found as the solution (Cardano's formula) of a 
cubic equation.

For $N=4$ the dynamical matrix has elements in terms of $\rho$, defined in
(\ref{defrho}):
\begin{eqnarray}
D_{n,n} &=& A + 2(C+D) + (3B + 16E) \rho^2 , \nonumber \\
D_{1,2} &=& D_{3,4} = -C - 2 E \rho^2 , \nonumber \\
D_{1,3} &=& D_{2,4} = -D(1 + e^{-iq}) , \\
D_{1,4} &=& ( -C -10 E \rho^2) e^{-iq}, \nonumber \\
D_{2,3} &=& -C -10 E \rho^2. \nonumber
\end{eqnarray}
The resulting secular equation is a quartic one and exact solutions for the
eigenvalues can be found using Ferrari's formula.

\begin{figure}
\begin{center}
\epsfig{file=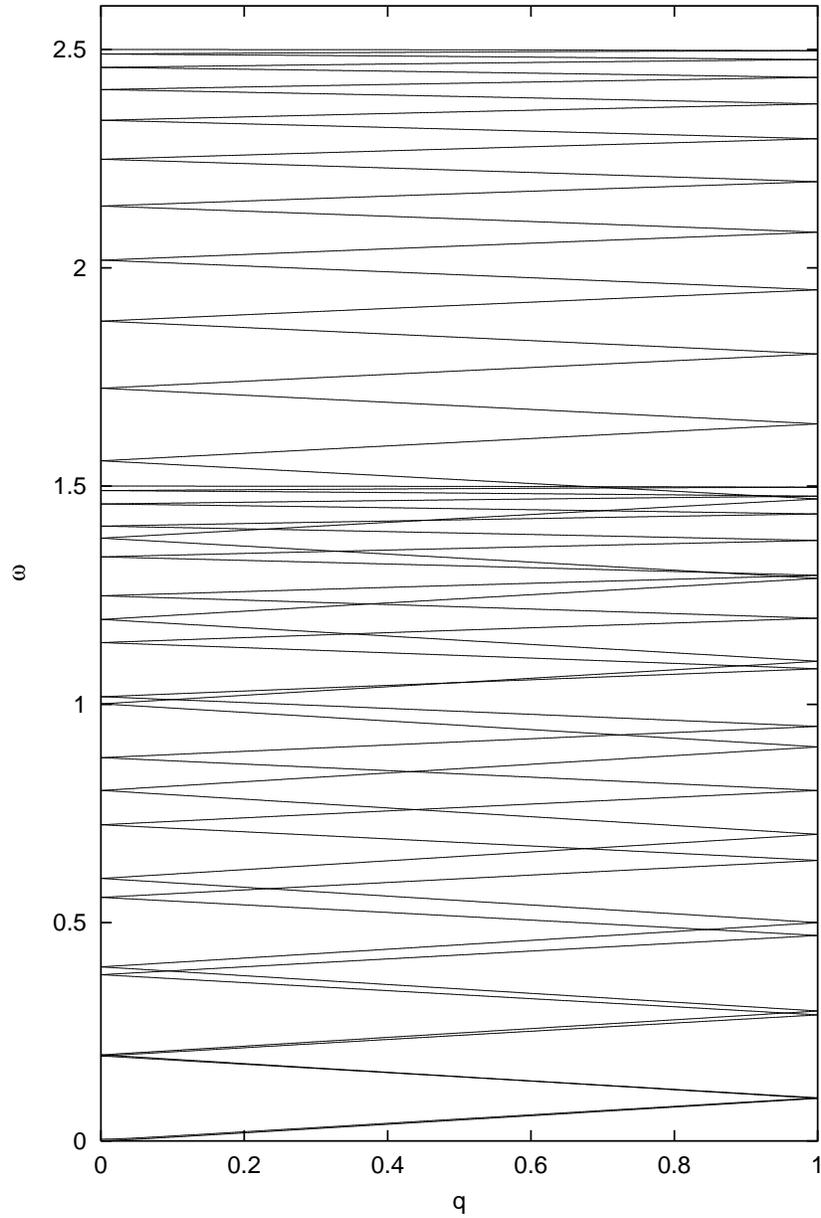,width=0.7\linewidth}
\end{center}
\caption{Phonon dispersion curves for $A=2.24999,~B=1,~C=1,~D=-1$ and 
$E=1$ in the 
commensurate approximation 62/13. This corresponds with the solution 
depicted in Figure
\ref{grdstate62-13}. $q$ is given in reduced units. There is
one branch with $\omega \rightarrow 0$ for $q \rightarrow 0$: the phason
branch. Just above this lies the amplitudon branch.}
\label{spec62-13}
\end{figure}

For solutions with larger periods we have to rely on numerical
calculations. As an example we again consider the ground state for
$A=2.24999,~B=1,~C=1,~D=-1,~E=1$. See also the end of Section III. The 
calculated phonon dispersion curves in 
the commensurate approximation $\lambda = \frac{62}{13}$ are given in Figure
\ref{spec62-13}.

\section{Calculation of phase diagrams}

In this section we present some phase diagrams, calculated partly
analytically, partly numerically. The traditional method to find the ground
state numerically is to solve the equations for equilibrium (\ref{statcon}).
However, these equations also hold for metastable states, maxima and
saddlepoints and it may happen
that one finds a metastable state instead of the true ground state. This
problem is not present for the so-called effective potential method
(EPM), introduced by Griffiths and Chou. \cite{Griff-Chou} This method in
principle always gives the ground state. Originally it was used to study
Frenkel-Kontorova and similar one-dimensional models with only
nearest-neighbor interaction. As an interesting application of this method,
we mention a study of the ground state of the chiral XY model in a field.
\cite{yokoi} Below we give a brief outline of the method.

Consider a one-dimensional system with only nearest-neighbor interaction in     its ground state. If one atom is displaced from its equilibrium position 
(we assume that $x_n$ denotes the displacement), the surrounding
atoms will change their positions in order to minimize the total energy.
This deformation will in general cost some energy. A function, called the
`effective potential', describes the net energy cost as a function of the
positions of the atoms. This effective potential achieves its minimum on
points of the ground state \cite{Mar-Hood} and rigorous mathematical
statements can be made. \cite{Chou-Duffin}
Numerical procedures to find solutions are
based on discretization of the range $x_n$ of the atomic positions. The
$x_n$ can now only adopt a finite number of values.
\cite{Chou-Griff,Griff-stat,Floria-Griff} 
For models with interactions up to next-nearest neighbors, as in the case
of the extended DIFFOUR model, the EPM can be adapted, which will be
discussed in Appendix B. The proofs of the existence of solutions for
models with next-nearest-neighbor interactions, both in
the continuous and discretized version, are rather long and will be given 
in a separate paper. \cite{koertetal}

Using the EPM and equation (\ref{statcon}) we calculated various phase 
diagrams. First we varied both $A$ and $E$ with the
other parameters fixed: $B=1$, $C=1$ and $D=-1$. The resulting phase
diagram is given in Figure \ref{phasediagea}. From the analysis in 
Appendix A we know that the phase boundary for the paraelectric state for 
$|C| < 4$ is given by $A = \frac{1}{4} C^2 - 2C + 4$.
For $C=1$ and $D=-1$ we find $A = 2 \frac{1}{4}$. At
this boundary we have a transition to an incommensurate state with
wave vector $q = \arccos(\frac{-C}{4D}) = \arccos(\frac{1}{4}) \approx
4.76679213$. This is the state we discussed at the end of Section III. 
We can clearly see the effect of the $E$-term: for $E=0$
further decreasing of $A$ leads to a transition to a commensurate state 
with period 4. For $E < 0$ this transition can be followed by transitions
to $N=3$ or $N=2$ commensurate states. For $E$ sufficiently positive, the 
wavelength of the 
ground state increases for decreasing $A$. Between the
commensurate states incommensurate ones can be found. 
By increasing $E$, the region between paraelectric and ferroelectric phases
shrinks. A positive $E$ term favors long-wavelength solutions, with the
ferroelectric state being the extreme limit $(q=0)$. 

Figure \ref{phasediagac0} gives the phase diagram found for $B=1$, $D=-1$,
$E=0$ and varying both $A$ and $C$. This is the phase diagram for the
original DIFFOUR model. \cite{tedalo} We have seen that
the phase boundary for the paraelectric phase for $|C| < 4$ is 
a parabola symmetric around $C = 0$. For $|C| \geq 4$
this boundary is given by $A + 2C -2 = -2 + 2|C|$, two straight
lines. The parabola and the lines meet at $|C| = 4$, and have equal
derivative at this point. Note the symmetry $C \leftrightarrow -C$, which
implies \cite{ishidiff} that the modulation wave vectors for the system
with $+C$ and $-C$ are related by
\begin{equation}
q_C + q_{-C} = \frac{1}{2}
\end{equation}
in units of $2 \pi$. At
$(C=4, A=0)$ the paraelectric phase, the ferroelectric phase and the 
incommensurate phase become equal. The lines separating the paraelectric
phase from the incommensurate phase and the incommensurate phase from the
ferroelectric phase have equal derivative at this point. In Landau theory 
(see Section VII) such a point would be
called a Lifshitz point. From the symmetry $C \leftrightarrow -C$ it is 
obvious that there is also a Lifshitz point at $(C=-4, A=16)$. At this
point the paraelectric phase, the antiferroelectric phase and the 
incommensurate phase become equal. 

Figure \ref{phasediagac1} gives the phase diagram for $B=1$, $D=-1$,
$E=1$ in terms of $A$ and $C$. The symmetry $C \leftrightarrow -C$ is
no longer present. However, the phase boundary of the paraelectric phase
is independent of $E$.
Also the wavelength of the phase emanating from this boundary is the same.
In particular the positions of the Lifshitz points and the derivates at
these points do not change. Note
the boundary of the antiferroelectric phase: starting from the Lifshitz
point and going down in the phase diagram, it initially bends to the right
and then returns to lower values of $C$.
Figures \ref{phasediagac0} and \ref{phasediagac1} have been obtained by
solving equation (\ref{statcon}) and comparing the energies of the
solutions.

\begin{figure}
\begin{center}
\epsfig{file=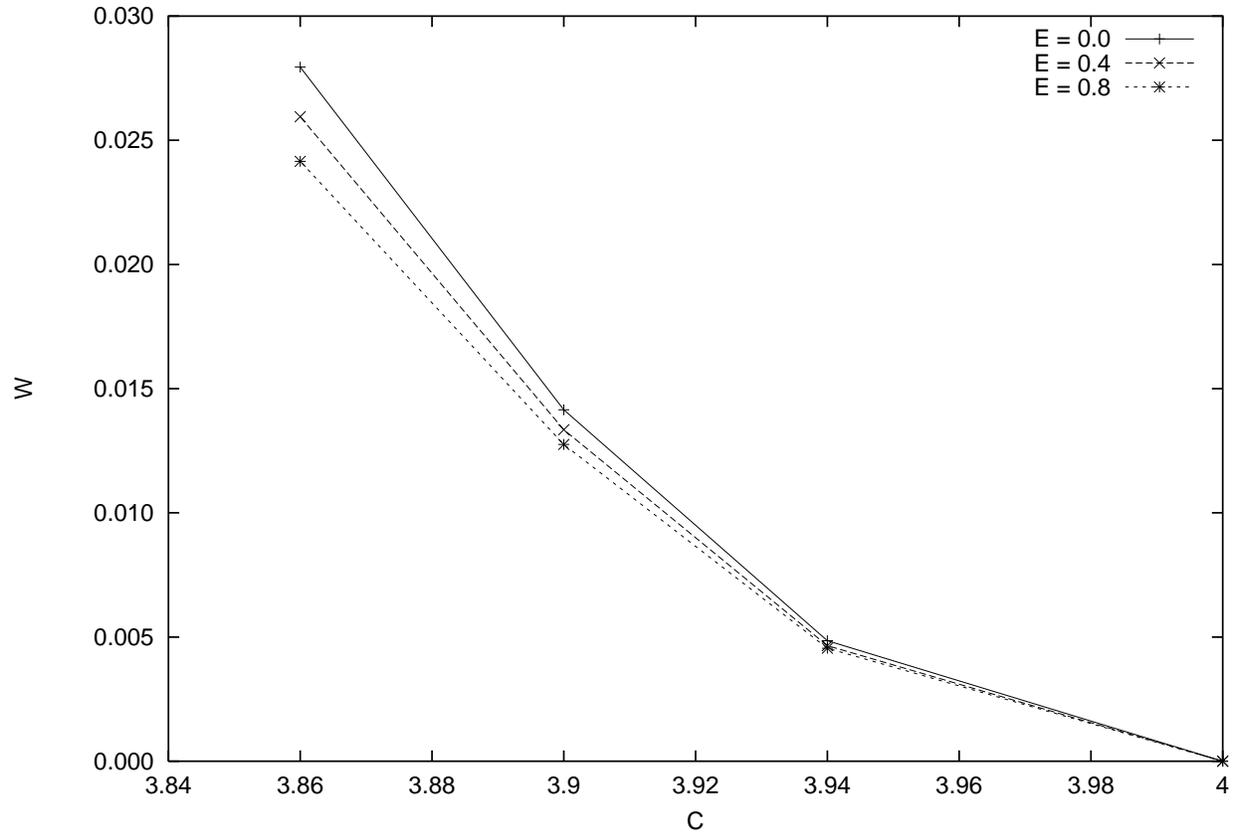,width=\linewidth}
\end{center}
\caption{Influence of the $E$ term on the wedge-width $W$ near the Lifshitz
point for $B=1$ and $D=-1$. Points are calculations, lines are drawn to
guide the eye.}
\label{wedge}
\end{figure}

We investigated the nearby surroundings of the Lifshitz point at $(C=4,
A=0)$ to look how the transition line from the ferroelectric phase to the
incommensurate phase changes by increasing $E$. See figure \ref{wedge} for
the results. One notices a tendency
towards a smaller wedge $W$ (the vertical distance between the 
paraelectric-incommensurate phase boundary and the 
incommensurate-ferroelectric phase boundary) by increasing $E$, as was to
be expected.

\section{Temperature dependent behavior}

As the model under consideration is one-dimensional with short-range
interactions, there is no phase transition possible at $T \neq 0$. If we 
however consider weakly interacting linear chains in a three dimensional
system, this system can be described by (\ref{edif}) as well if we
interpret the variables $x_n$ as averages over planes perpendicular to a
fixed direction (the $c$-axis). Phase transitions become possible due to
inter-chain couplings.

To study the temperature dependence of the parameters we take the
thermal average of the conditions for equilibrium (\ref{statcon}), 
resulting in
\begin{eqnarray}
&&A \langle x_n \rangle + B \langle x_n^3
\rangle + C \left( 2 \langle x_n \rangle - \langle x_{n-1} \rangle
- \langle x_{n+1} \rangle \right) + D \left( 2 \langle x_n \rangle -
\langle x_{n-2} \rangle - \langle x_{n+2} \rangle \right) \nonumber\\
&&+ E \left[ 4 \langle x_n^3 \rangle - 3 \langle x_n^2 x_{n-1} \rangle - 3
\langle x_n^2 x_{n+1} \rangle  + 2 \langle x_n x_{n-1}^2 \rangle + 2 \langle
x_n x_{n+1}^2 \rangle  - \langle x_{n-1}^3 \rangle - \langle x_{n+1}^3
\rangle \right] = 0
\label{therstat}
\end{eqnarray}
We have to distinguish between ground states with $\{\bar{x}_n\} = 0$ and 
ground states with $\{\bar{x}_n\} \neq 0$, where the $\{\bar{x}_n\}$ are 
solutions of (\ref{statcon}).

In the former case
we assume that the thermal fluctuations of the displacement $x_n$ 
do not depend on the lattice site, $\langle x_n^2 \rangle - \langle x_n
\rangle^2 \approx \langle x_m^2 \rangle - \langle x_m \rangle^2$, and
if we furthermore approximate the correlations by $\langle x_n^2 x_m 
\rangle \approx \langle x_n^2 \rangle \langle x_m \rangle$, 
the following holds:
\begin{eqnarray}
&& E \left[  - 3 \langle x_n^2 x_{n-1} \rangle - 3 \langle x_n^2 x_{n+1} 
\rangle + 2 \langle x_n x_{n-1}^2 \rangle + 2 \langle
x_n x_{n+1}^2 \rangle - \langle x_{n-1}^3 \rangle - \langle x_{n+1}^3
\rangle \right] \nonumber\\
&& + (B+4E) \langle x_n^3 \rangle \nonumber \\
&\approx&
E \left[ - 3 \langle x_n \rangle^2 \left( \langle x_{n-1}
\rangle + \langle x_{n+1} \rangle \right) + 2 \langle x_n \rangle
\left( \langle x_{n-1} \rangle^2 + \langle x_{n+1} \rangle^2 \right)
- \left( \langle x_{n-1} \rangle^3 + \langle x_{n+1} \rangle^3 \right)
\right] \nonumber \\
&& + 4 E \left( \langle x_n^2 \rangle - \langle x_n \rangle^2 \right) \left( 2
\langle x_n \rangle - \langle x_{n-1} \rangle - \langle x_{n+1} \rangle
\right) \nonumber \\
&&+(B+4E) \langle x_n \rangle^3 + B \left( \langle x_n^2 \rangle - \langle 
x_n \rangle^2 \right) \langle x_n \rangle.
\end{eqnarray}
Inserting the last expression in (\ref{therstat}) we see that
the conditions for equilibrium for the thermal average of the displacement,
$\langle x_n \rangle$, have the same form as those for the displacements
$x_n$ themselves; the only difference being the replacement of the
parameters $A$ and $C$ by temperature dependent ones:
\begin{eqnarray}
A &\rightarrow& A + B  T, \nonumber \\
C &\rightarrow& C + 4 E  T,
\label{part}
\end{eqnarray}
where $T = \langle x_n^2\rangle - \langle x_n \rangle^2$ is a measure   
of the thermal fluctuations. So a change in temperature will renormalize 
both parameters $A$ and $C$ (unlike in the DIFFOUR model with $E = 0$).

\begin{figure}
\begin{center}
\epsfig{file=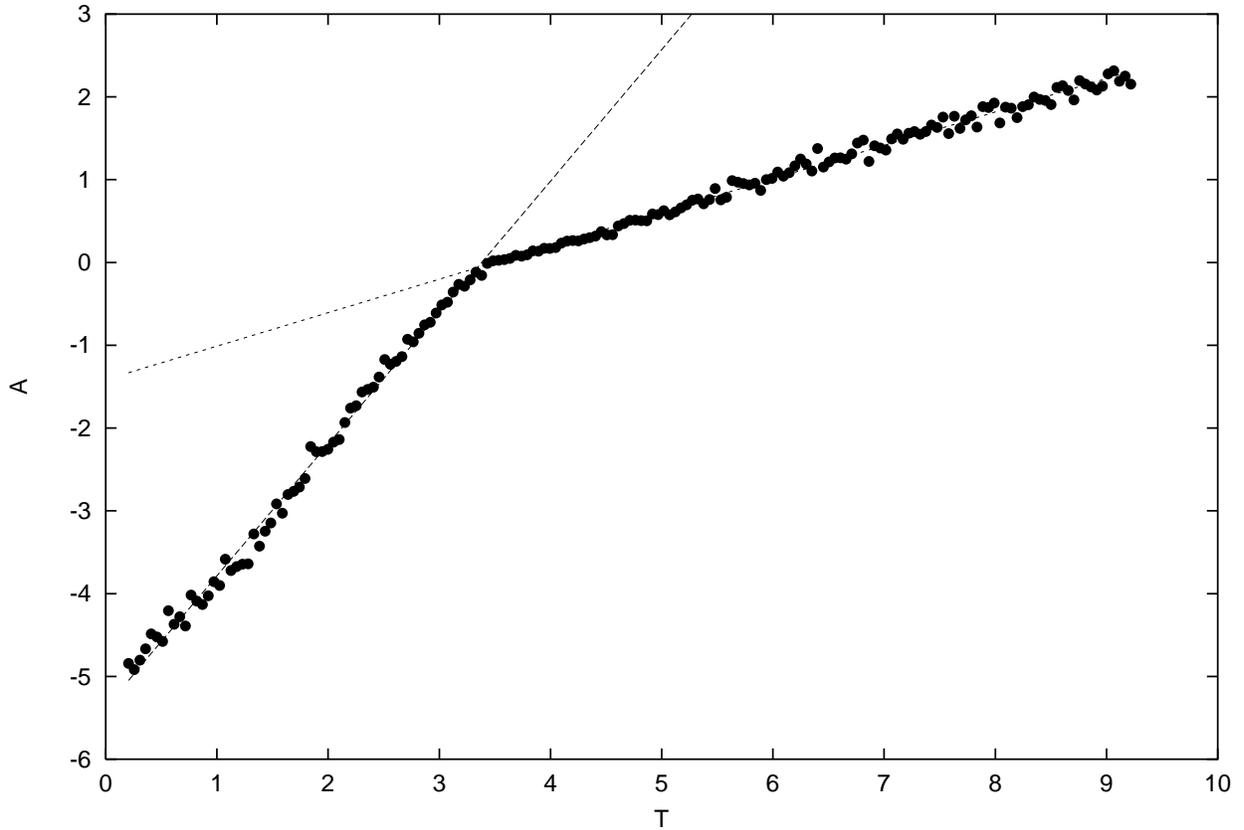,width=\linewidth}
\end{center}
\caption{Temperature dependence $A(T)$ calculated on a 3D version of 
(\ref{edif}) with harmonic nearest-neighbor coupling and $A(0) = -5$, $B
= 5$, $D=E=0$. Points are calculations, \protect \cite{alexpre} lines are 
fits with a linear function. The phase transition takes place at $A = 0$.}
\label{alexeytemp}
\end{figure}

For all other ground state solutions ($\bar{x}_n \neq 0$) we
calculate the thermal averages around $\bar{x}_n$, where the $\{\bar{x}_n
\}$ satisfy (\ref{statcon}),
\begin{equation}
\langle x_n^p x_m^q \rangle = \frac{\int \int x_n^p x_m^q e^{- \beta(x_n -
\bar{x}_n)^2/2} e^{- \beta(x_m - \bar{x}_m)^2/2} \text{d}x_n
\text{d}x_m}{\int \int e^{- \beta(x_n - \bar{x}_n)^2/2} e^{- \beta(x_m -
\bar{x}_m)^2/2} \text{d}x_n \text{d}x_m} = \langle x_n^p \rangle
\langle x_m^q \rangle,
\end{equation}
where $\beta = 1/T$. Three different integrals have to be calculated,
yielding
\begin{eqnarray}
\langle x_n^3 \rangle &=& \bar{x}_n^3 + 3 \bar{x}_n T \nonumber \\
\langle x_n^2 \rangle &=& \bar{x}_n^2 + T \\
\langle x_n \rangle &=& \bar{x}_n \nonumber
\end{eqnarray}
Substitution in (\ref{therstat}) and comparison with (\ref{statcon}) then
leads to
\begin{eqnarray}
A &\rightarrow& A + (3B + 4E)  T, \nonumber \\
C &\rightarrow& C + 6E  T.
\label{fert}
\end{eqnarray}
Note that the parameter $E$ now also enters in the temperature dependence
of $A$. This linear behavior in $T$, with a kink at the temperature where
the transition from the paraelectric phase to the incommensurate or the
ferroelectric phase takes place, is corroborated by Monte Carlo
calculations. \cite{radescu,alexpre} Some of the results \cite{alexpre} are     shown in
Figure \ref{alexeytemp}. The results (\ref{part}) and (\ref{fert}) are in
sharp contrast with the assumptions made in standard Landau theory, to be 
discussed in Section VII, that there is only one temperature dependent 
parameter, and that its behavior above and below the transition temperature     is the same.

It is now straightforward to calculate the temperature dependent  
ground states and stability limits by making substitutions (\ref{part}) and
(\ref{fert}). We especially would like to focus on the phonon branches in 
the paraelectric and ferroelectric phases. Of experimental interest is the
ratio between the slopes of the soft phonon mode in the ferroelectric and
paraelectric phase, the so-called $R$-parameter:
\begin{equation}
R = \frac{ {\rm d} \omega^2 / {\rm d} T|_{\rm ferro}}{ {\rm d}
\omega^2 / {\rm d} T|_{\rm para}}.
\end{equation}
Self-consistent renormalized phonon theory 
\cite{bruce} gives the result $R = -2$. In experiments \cite{stephans6}
however very often $R \neq -2$ is found. Taking into account the 
temperature dependence given in (\ref{part}) and (\ref{fert}), we find
using (\ref{parphonon}) and (\ref{ferphonon}) 
\begin{equation}
R = \frac{-2(3B+4E) - 32 E^2 / B \sin^2(q/2)}{B + 16E \sin^2(q/2)},
\end{equation}
which, for $E=0$, gives (at the center of the Brillouin zone) $R = -6$ 
instead of $R = -2$, and for $E \neq 0$ can take on any value as long as $B
+ 16 E > 0$ is satisfied.
Molecular dynamics simulations on a 3-dimensional $\phi^4$ lattice by 
Padlewski {\it et al.} \cite{padlewski} show that $R=-2$ holds only for 
systems with long-range couplings being in the displacive limit. However,
Sollich {\it et al.} \cite{sollich} showed that this double limit of
displaciveness and long range interaction is not necessary if the system is
displacive enough: they found $R = -2$ for a system with only
nearest-neighbor interactions, thereby questioning Padlewski's claim
\cite{padlewski} of having studied a system in the displacive limit.

\section{Comparison with the continuum theory}

The continuum limit of the extended DIFFOUR model leads to a well-known
expansion. By replacing the differences in the general order parameter $x_n$
by differentials in the principal order parameter for ferroelectrics,
the polarization $P(z)$,
\begin{eqnarray}
(x_n - x_{n-1})^2 &\rightarrow& \left( \frac{{\rm d} P}{{\rm d} z}
\right)^2, \nonumber \\
(x_{n-1} + x_{n+1} - 2 x_n)^2 &\rightarrow& \left( \frac{{\rm d}^2 P}
{{\rm d} z^2} \right)^2, 
\end{eqnarray}
and rearranging terms we arrive at the free energy density
\begin{equation}
f =  \frac{\alpha}{2} P^2 + \frac{\beta}{4} P^4 +
\frac{\kappa}{2} \left( \frac{{\rm d} P}{{\rm d} z} \right)^2 + 
\frac{\lambda}{2} \left( \frac{{\rm d}^2 P}{{\rm d} z^2} \right)^2 + 
\frac{\eta}{2} P^2 \left( \frac{{\rm d} P}{{\rm d} z} \right)^2, 
\label{endens}
\end{equation}
with $\alpha = A$, $\beta = B$, $\kappa = C + 4D$, $\lambda = -D$ and $\eta
= 2E$. The above free energy density was used by Ishibashi and Shiba
\cite{ishibashi} to study phase transitions in NaNO$_2$ and SC(NH$_2$)$_2$ 
(thiourea), proper ferroelectrics in which the polarization component of
interest transforms according to a one-dimensional irreducible 
representation. The $\eta$-term is allowed by symmetry because it is the
product of the two invariants $P^2$ and $({\rm d}^2 P / {\rm d} z^2)^2$.
Alternatively, as both sodium nitrite and thiourea admit an interaction of
$P$ with another mode $u$ (strain for example), Dvo\v r\'ak \cite{dvorak}
showed that the $\eta$-term accounts in an effective manner for this
interaction, thereby reducing $g(P,u) \rightarrow f(P)$.

By taking the Fourier transform of the above free energy density we find
\begin{equation}
\tilde{f} = \left( \frac{\alpha}{2} + \frac{\kappa}{2} q^2 +
\frac{\lambda}{2} q^4 \right) P_q^2 + \left( \frac{\beta}{4} +
\frac{\eta}{2} q^2 \right) P_q^4.
\end{equation}
This justifies the choice of the $E$-term in the extended version of the 
DIFFOUR model, discussed in Section II.
A term $\propto ({\rm d} P / {\rm d} z)^4$ has been included in
the free energy expansion by Jacobs {\it et al.}, \cite{jacobsetal} but by
taking the Fourier transform one finds $\propto q^4 P_q^4$ which is of
higher order than the $\eta$-term used here.

In a seminal paper Hornreich {\it et al.} \cite{hornreich1} discussed a
multicritical point of a new type, which they called a Lifshitz point. In
the spherical model limit they were able to calculate critical exponents
and the shape of the phase boundaries of $2^{\rm nd}$ order and
$1^{\rm st}$ order transitions in the vicinity of the Lifshitz point.
\cite{hornreich2} Let us return to the above free energy density to give a
definition \cite{selke} of the Lifshitz point. At an ordinary paraelectric 
to ferroelectric phase transition the coefficient $\alpha$ changes sign. If     we have an additional incommensurate phase we need the $\kappa$ and
$\lambda$-terms, and at the Lifshitz point $\kappa = 0$. Higher-order
terms in the expansion are needed for stabilization. Converting $\alpha =
0, \kappa = 0$ to variables in the extended DIFFOUR model we find $A=0,C=4$
(for $D=-1$). This is exactly the position of the Lifshitz point found in
Section V. There is another analogy between Landau theory and the DIFFOUR
model: Michelson \cite{michelson} showed
that for systems with uniaxial polarization the phase transition lines
separating the paraelectric phase from the incommensurate phase and the
incommensurate phase from the ferroelectric phase are tangent at the
Lifshitz point. This feature is also present in figures \ref{phasediagac0}
and \ref{phasediagac1}.

Let us now discuss some properties of the solutions found in Landau theory.
Ground states minimize the total free energy
\begin{equation}
F = \frac{1}{d} \int_0^d f(z) {\rm d}z, 
\end{equation}
and can be found by solving the Euler-Lagrange equation
\begin{equation}
\lambda \frac{{\rm d}^4 P}{{\rm d} z^4} - \kappa \frac{{\rm d}^2 P}
{{\rm d} z^2} - \eta \left[ P \left( \frac{{\rm d} P}{{\rm d} z} \right)^2
+ P^2 \frac{{\rm d}^2 P}{{\rm d} z^2} \right] + \alpha P + \beta P^3 = 0.
\label{eulerlag}
\end{equation}
Golovko \cite{golovko} was able to obtain exact solutions for some special
values of the parameters in a slightly more general free energy density
(he added a term $\frac{\gamma}{6} P^6$ to the expansion (\ref{endens})). 
However, his method is not
general and we will not discuss it further. Instead we follow a different
approach: numerically solving \cite{ishibashi} equation (\ref{eulerlag}) 
shows that the solutions contain practically only one harmonic, the
amplitude of higher harmonics is at most 3.5\% of the former.

As usual in Landau theory only the coefficient $\alpha$ is 
temperature dependent: $\alpha = \alpha_0(T - T_c)$. It is found that
between the high-temperature paraelectric solution 
\begin{equation}
P(z) = 0, \qquad F = 0,
\end{equation}
and the low-temperature ferroelectric solution
\begin{equation}
P(z) = \sqrt{- \frac{\alpha}{\beta}}, \qquad F = - \frac{\alpha^2}{4 \beta}, 
\end{equation}
an incommensurate solution exists. Just below the paraelectric-incommensurate   transition at $\alpha_i = \alpha_0(T_i - T_c) = \frac{\kappa^2}{4 \lambda}$ it  has the form \cite{golovko}
\begin{equation}
P(z) = \rho_0 \cos(qz), \qquad F = - \frac{(\alpha_0 - \alpha)^2}{2 \left(3
\beta + 2 \eta q_0^2 \right)}.
\end{equation}
The amplitude $\rho_0$ and wave vector $q$ are given by
\begin{eqnarray}
\rho_0^2 &=& \frac{4(\alpha_0 - \alpha)}{3 \beta + 2 \eta q_0^2}, \nonumber
\\
q &=& q_0 \left( 1 + \frac{\eta}{8 \kappa} \rho_0^2 \right), \\
q_0^2 &=& - \frac{\kappa}{2 \lambda}. \nonumber
\end{eqnarray}
The $\eta$-term makes the incommensurate phase less stable when $\eta$ is
positive, implying that the transition temperature from the incommensurate 
state to the ferroelectric state increases as $\eta$ increases. See also
the discussion by Tol\'edano. \cite{toledano} In the discrete model a 
positive $E$-term is responsible for this effect.

\section{Conclusions and outlook}

In this paper we have calculated various properties of an 
extension of the DIFFOUR model. For this purpose a next-nearest-neighbor
generalization of the Effective Potential Method was developed. 
The shape of the paraelectric phase boundary was 
proven rigorously, elaborating on a former proof which only included 
nearest-neighbor interactions. We found that the phase diagram changes 
considerably due to the extra $E$-term, but the transition at the
paraelectric phase boundary does not depend on $E$. Positive $E$ favors
longer-period solutions.

By taking thermal fluctuations in two different regimes into account the
parameters $A$ and $C$ can be considered as effectively temperature
dependent. For $C$ this holds only for nonzero $E$, which explains the
relevance of this extra term. This has
strong consequences for two experimentally easy accessible quantities: the 
temperature dependence of the modulation wave vector, and the ratio between
the slopes of the soft phonon mode in the ferroelectric and paraelectric
phases ($R$-parameter).

Although lattice and continuum models have some features in common, the 
differences are more striking. A lattice model would be a more natural
choice than the phenomenological Landau treatment of incommensurate phases.
Discrete models do not need {\it ad hoc} lock-in terms to explain different     commensurate and incommensurate phases. Complex phase 
diagrams can in principle be obtained using a simple Hamiltonian which
takes into account the discreteness of a lattice.

The Sn$_2$P$_2$(S$_{1-x}$Se$_x$)$_6$ crystal family seems to be an
excellent system for our future research: it is uniaxial, has an
exceptional Lifshitz point in the composition-temperature phase diagram,
shows cross-over effects from order-disorder to a displacive type of phase
transition, and displays an interesting modulation wave vector behavior. All
these phenomena can in principle be explained by the extended DIFFOUR
model.

\begin{acknowledgments}
We thank Stephan Eijt for drawing our attention to this problem and Alexey
Rubtsov for providing us with the results of the Monte Carlo calculations.
\end{acknowledgments}

\appendix

\section{Exact results for paraelectric and ferroelectric phases}

In this Appendix explicitly calculated phase boundaries are given for the
extended DIFFOUR model.
We first consider $E=0$ and then discuss the effect of $E \neq
0$. Let us start with writing $V$ in the form
\begin{equation}
V = \sum_{n} \left\{ \frac{a}{2} x_n^2 + \frac{1}{4} x_n^4 + cx_n x_{n-1}
+ dx_n x_{n-2} \right\},
\end{equation}
with $d=\pm 1$. The remaining parameters $a,c,d$ are the tilde parameters
defined in (\ref{deftilde}) after normalization of $\tilde{B}$ and 
$\tilde{D}$. Let us now try to write this as 
\begin{equation}
V=\sum_n \left\{ p(x_n-qx_{n-1}-rx_{n-2})^2+\frac{1}{4}x_n^4 \right\}.
\label{A2}
\end{equation}
Comparison of the two expressions yields
\begin{eqnarray}
p(1+q^2+r^2) &=& \frac{a}{2}, \nonumber \\
p(-2q+2qr) &=& c, \\
-2pr &=& d. \nonumber
\end{eqnarray}
From this one can see that if it is possible to write the potential in
this form and $a$ is positive, then $p$ is positive. Eliminating $q$ and
$r$ from the
above equations yields the following fourth order polynomial equation
(assuming non-zero $a,c$ and $d$):
\begin{equation}\label{poly-4}
16p^4+(16d-8a)p^3+(8d^2+4c^2-8ad)p^2+(4d^3-2ad^2)p+d^4=0. 
\end{equation}

First consider the $d=+1$ case. Equation (\ref{poly-4}) then has the 
following complex solutions
\begin{eqnarray}
p&=&\frac{1}{8}\Bigg(-2+a+\sqrt{(2+a)^2-4c^2} \nonumber\\
&&\pm \sqrt{2}\sqrt{-4+a^2-2c^2+(-2+a)\sqrt{(2+a)^2-4c^2}}\Bigg), 
\label{A5}
\end{eqnarray}
\begin{eqnarray}
p&=&\frac{1}{8}\Bigg(-2+a-\sqrt{(2+a)^2-4c^2} \nonumber\\
&&\pm \sqrt{2}\sqrt{-4+a^2-2c^2-(-2+a)\sqrt{(2+a)^2-4c^2}}\Bigg).
\end{eqnarray}
The first requirement for having a real solution is that
$(2+a)^2-4c^2\geq 0$, i.e. $a\geq -2+2\vert c\vert$.
Consider the first two solutions (\ref{A5}).
First look at $\vert c \vert \geq 4$. Then for $a\geq
-2+2\vert c\vert$ we find $(-2+a)\sqrt{(2+a)^2-4c^2}>0$. And for the first     term in the root we find $-4+a^2-2c^2>2c^2-8\vert c\vert \geq 0$. 
So for $\vert c \vert
\geq 4$ the only requirement for having a real (positive, $a>0$) 
solution is $a\geq -2+2\vert c\vert$. For $\vert c\vert
<4$ we have $-4+a^2-2c^2+(-2+a)\sqrt{(2+a)^2-4c^2}=0$ for $a=2+\frac{1}{4}
c^2$. It can be seen that the argument of the root
is positive for $a>2+\frac{1}{4}c^2$. So, for $\vert c \vert <4$ the
requirement for having a real (positive, $a>0$) solution is: $a\geq
2+\frac{1}{4}c^2$ (then $a>-2+2\vert c \vert$ automatically holds too).

The requirements for $\vert c \vert \geq 4$ and $\vert c \vert <4$ form a
continuous line in the $a-c$-parameter space. Above this line $V$ can
be written as (\ref{A2})
with $p$ positive. Therefore $V\geq 0$. The lower bound is reached by
the trivial solution which always exists, so the paraelectric phase is the
ground state above this line. In Sec. IV it is shown that the above
line corresponds exactly with the stability lines of the trivial solution,
showing that the modulated phases arise from the destabilization of
the trivial solution due to the condensation of a soft phonon mode. 

In case $d = -1$ the solutions of the fourth order polynomial equation are
\begin{eqnarray}
p&=&\frac{1}{8}\Bigg(2+a+\sqrt{(-2+a)^2-4c^2} \nonumber\\
&&\pm \sqrt{2}\sqrt{-4+a^2-2c^2+(2+a)\sqrt{(-2+a)^2-4c^2}}\Bigg),
\label{A7}
\end{eqnarray}
\begin{eqnarray}
p&=&\frac{1}{8}\Bigg(2+a-\sqrt{(-2+a)^2-4c^2} \nonumber\\
&&\pm \sqrt{2}\sqrt{-4+a^2-2c^2-(2+a)\sqrt{(-2+a)^2-4c^2}}\Bigg).
\end{eqnarray}
Look at the two solutions (\ref{A7}).
The first condition is  $a\geq 2+2\vert c \vert$. 
$(2+a)$ is positive if this
requirement is fulfilled. Further $-4+a^2-2c^2\geq 2c^2+8\vert c
\vert \geq 0$. So, here we have only one requirement for all $c$,
namely $a\geq 2+2\vert c \vert$. Above this line the paraelectric phase is 
the ground state.

With the same sort of reasoning we can also try to prove that for
certain parameter values the ground state is ferroelectric
Here we work with the following form of $V$ (with $D=\pm1$):
\begin{equation}
V=\sum_n \left\{ \frac{A}{2}x_n^2+\frac{1}{4}x_n^4+\frac{C}{2}
(x_n-x_{n-1})^2+\frac{D}{2}(x_n-x_{n-2})^2 \right\}.
\end{equation}
We try to write this as 
\begin{equation}
V=\sum_n \left\{ \frac{A}{2}x_n^2+\frac{1}{4}x_n^4+P(x_n-Qx_{n-1}-
Rx_{n-2})^2 \right\}.
\label{A10}
\end{equation}
Comparison of the two expressions yields
\begin{eqnarray}
P(1+Q^2+R^2) &=& C+D, \nonumber \\
P(-2Q+2QR) &=& -C, \\
-2PR &=& -D. \nonumber
\end{eqnarray}
If there exists a solution and $C+D$ is positive, then $P$ is positive.
Rewriting the above equations yields the following fourth order
polynomial equation (assuming nonzero $C$ and $D$):
\begin{equation}
16P^4+(-32D-16C)P^3+(24D^2+16CD+4C^2)P^2+(-8D^3-4CD^2)P+D^4=0.
\end{equation}

For the case $D=-1$ the complex solutions are
\begin{equation}
P=\frac{1}{4} \left(-2+C\pm \sqrt{C}\sqrt{-4+C} \right),
\end{equation}
both having multiplicity 2. For $0<C<4$ the solution is not real.
For $C\geq 4$ the solution is real. In order to have a positive
solution we must have $C+D>0$, so $C>1$. So, for $C>4$ the potential
can be written as (\ref{A10}) with $P$ positive. So
\begin{equation}
V\geq \sum_n \left\{ \frac{A}{2}x_n^2+\frac{1}{4}x_n^4 \right\}.
\end{equation}
The ferroelectric phase, which exists if $A<0$, reaches this lower bound. So
for $B=1, D=-1, A<0$ the ground state is ferroelectric for $C>4$. 
In terms of the tilde parameters: for $\tilde{B}=1, \tilde{D}=1,
\tilde{A}<-2-2\tilde{C}$, the ferroelectric phase is the ground state for
$\tilde{C}<-4$. 

For $D=+1$ the complex solutions are
\begin{equation}
P=\frac{1}{4} \left(2+C\pm \sqrt{C}\sqrt{4+C} \right).
\end{equation}
For $-4<C<0$ the solution is not real. For $C\geq 0$ the solution is
real. In that case $P$ is positive. For $B=1, D=1, A<0$ the ferroelectric
phase is the ground state for $C\geq 0$. In other words: for
$\tilde{B}=1, \tilde{D}=-1, \tilde{A}<2-2\tilde{C}$ the ferroelectric phase
is the ground state for $\tilde{C}\leq 0$. 
In terms of the tilde parameters analogous statements about the
anti-ferroelectric phase can easily be made ($\tilde{C}\leftrightarrow 
-\tilde{C}$).

In case $E \neq 0$ the following holds: the parts of the phase diagram
where the paraelectric phase is
the ground state in the DIFFOUR model $(E=0)$ also belong to the 
paraelectric phase
for this extended model for all allowed values of $E$. In this extended
model there are no other parts of the phase diagram where the trivial
solution is the ground state, because the stability conditions for this
solution are the same as in the DIFFOUR model (see Section IV).

For the ferroelectric phase the statements are less rigorous: if the 
ferroelectric phase is
the ground state in the DIFFOUR model $(E=0)$, then it is also the ground
state in the extended model for $E>0$. Again we can prove that $V \geq
\sum_n \left\{ \frac{A}{2} x_n^2 + \frac{B}{4} x_n^4 \right\}$. However,
for positive $E$ the part of the phase diagram where the ferroelectric
phase is the ground state becomes bigger.

\section{Effective Potential Method for next-nearest neighbors}

In this Appendix, which is based on the account given by Griffiths,
\cite{Griff-stat} we discuss how the EPM can be generalized to be applicable
to systems in which there is next-nearest neighbor interaction. 
Consider a classical one-dimensional chain of atoms. The
total potential energy of the system is given by
\begin{equation}\label{eq:sec_neigh_pot}
H=\sum_{n=-\infty}^{\infty} \left\{ V(x_n)+W(x_{n+1},x_n)+D(x_{n+2},x_n)
\right\}.
\end{equation}
So, interactions up to next-nearest-neighbors are included. 
The effective potential method that is used to find ground states for
systems with
interaction with first neighbors, can also be used for 
systems where second neighbor interaction is included. 
Instead of a scalar variable at site $n$, one now has to deal with a vector
consisting of the values for $x$ for two adjacent atoms. \cite{Chou-Griff} 
Writing ${\bf x}_n = (x_{2n},x_{2n-1})$ 
the above potential energy is of the form
\begin{equation}
H=\sum_{n}  K({\bf x}_n, {\bf x}_{n-1}) ,
\end{equation}
with 
\begin{eqnarray}
K({\bf x}_n, {\bf x}_{n-1}) &=& V(x_{2n})+V(x_{2n-1})+W(x_{2n},x_{2n-1})
+W(x_{2n-1},x_{2n-2}) \nonumber\\
&&+D(x_{2n},x_{2n-2})+D(x_{2n-1},x_{2n-3}).
\end{eqnarray}
The fact that here a vector at site $n$ is considered does not change
the EPM and the proofs given for this method.
\cite{Griff-Chou,Chou-Griff,Griff-stat,Floria-Griff}
The solution can be found by solving
\begin{equation}
\eta+R({\bf x}_{n})=\min_{{\bf x}_{n-1}} \left\{R({\bf x}_{n-1}) +
K({\bf x}_{n}, {\bf x}_{n-1}) \right\}.
\end{equation}
Here $\eta$ is 2 times the ground state energy per particle.
The vector consists of two components with respect to which one has to
minimize. Because of this minimization over two components, which has to be
performed frequently, the numerical procedures based on discretization of
the range of possible $x_n$-values will take a very long time.
The method we used is slightly different. Consider the $n^{\rm th}$ couple of
adjacent atoms. 
Couple $(n-1)$ does not consist of two other atoms as is the case above.
Instead, the right atom of couple $(n-1)$ is the same as the
left atom of couple $n$. In this case only a minimization over
one atomic degree of freedom (the `position') is left. This leads to more 
reasonable computation
times. The fact that couple $(n-1)$ is not independent of couple $n$ 
requires an adaptation of the proof of the existence of a
solution both in the continuous case and in the discretized case
used for numerical procedures. We only give an outline for the method,
proofs of the existence of solutions and generalization to systems with
interactions up to $s^{\rm th}$ neighbors are to be given in a separate
paper. \cite{koertetal} The method 
for deriving the equations and the numerical procedures \cite{Floria-Griff}
for solving the equations remain essentially the same. Numerical
calculations suggest that the error in $\eta$ has a cubic dependence on the
grid size rather than a quadratic dependence for the Frenkel-Kontorova
model. \cite{Chou-Griff}

Let us give the following explanation \cite{Mar-Hood} for the method: 
Imagine that a system described by (\ref{eq:sec_neigh_pot}) is in its
ground state. If we now change the positions of two adjacent atoms, the
surrounding atoms will in general also change their positions in order
to minimize the total energy. This net energy change caused by the
deformation of one couple will be called the effective
two-particle-potential. This will describe the energy cost as a function 
of the positions of two adjacent atoms. 
At site $n$, the effective two-particle-potential $R(x_{n+1},x_n)$,
due to the presence of the atoms $i<n$, can be formally written as
\begin{equation} \label{eq:def_right_eff_ener}
R(x_{n+1},x_n)\equiv \min_{i<n} \left\{ \sum_{i\leq n+1} \left[ V(x_i)+
W(x_i,x_{i-1})+D(x_i,x_{i-2})-\eta \right] \right\}, 
\end{equation}
where the minimum is taken over all atomic positions $x_i$ with $i<n$
and $\eta$ is the (unknown) ground state energy per particle.
By rewriting this equation, one obtains 
\begin{eqnarray}
R(x_{n+1},x_n)= \min_{x_{n-1}}  \min_{i<{n-1}}
\Bigg\{ && \sum_{i\leq n} \left[ V(x_i)+W(x_i,x_{i-1})+D(x_i,x_{i-2})-\eta
\right] \nonumber\\
&&+V(x_{n+1})+W(x_{n+1},x_n)+D(x_{n+1},x_{n-1})-\eta \Bigg\},
\end{eqnarray}
which gives
\begin{equation}
\eta +R(x_{n+1},x_n)=V(x_{n+1})+W(x_{n+1},x_n)+
\min_{x_{n-1}} \left[ R(x_n,x_{n-1})+D(x_{n+1},x_{n-1}) \right].
\label{B7}
\end{equation}
This is the minimization eigenvalue equation for $R$.
The same procedure can be followed for the effect of the atoms
\mbox{$i>{n+1}$}.
The effective two-particle-potential due to these atoms is called
$S(x_{n+1},x_n)$, which gives
\begin{equation}
\eta+S(x_{n+1},x_n)=V(x_n)+W(x_{n+1},x_n)+\min_{x_{n+2}} \left[
S(x_{n+2},x_{n+1})+ D(x_{n+2},x_n) \right].
\label{B8}
\end{equation}
The total effective two-particle-potential $F(x_{n+1},x_n)$ of a couple of
adjacent atoms in
a double infinite chain, is given by
\begin{equation}
F(x_{n+1},x_n)=R(x_{n+1},x_n)+S(x_{n+1},x_n)-V(x_{n+1})-V(x_n)-W(x_{n+1},x_n),
\end{equation}
where the last three terms are subtracted on the right side to avoid
double counting.

The equations (\ref{B7}) and (\ref{B8}) can also be obtained in another way. 
\cite{Chou-Griff,Griff-stat}
Let $\tilde{R}_N(x_{n+1},x_n)$ be the minimal energy of a chain of $N$
atoms with the constraint that the atoms $N$ and $N-1$ are at
fixed positions $x_{n+1}$ and $x_n$ respectively, while the other
atoms are free to rearrange themselves in an optimal way so as to
minimize the total energy. This leads to
\begin{eqnarray}
\tilde{R}_2(x_{n+1},x_n) &=& V(x_{n+1})+V(x_n)+W(x_{n+1},x_n), \\
\tilde{R}_3(x_{n+1},x_n) &=& V(x_{n+1})+V(x_n)+W(x_{n+1},x_n)
\nonumber\\
&& +\min_{x_{n-1}} \left[ V(x_{n-1})+W(x_n,x_{n-1})+D(x_{n+1},x_{n-1})
\right] \nonumber\\
&=&V(x_{n+1})+W(x_{n+1},x_n)+\min_{x_{n-1}} \left[\tilde{R}_2(x_n,x_{n-1})+
D(x_{n+1},x_{n-1}) \right], \\
\tilde{R}_{N+1}(x_{n+1},x_n) &=&V(x_{n+1})+W(x_{n+1},x_n)+\min_{x_{n-1}}
\left[ \tilde{R}_N(x_n,x_{n-1})+D(x_{n+1},x_{n-1}) \right].
\end{eqnarray}
Now assume that for $N\rightarrow \infty$, $\tilde{R}_N(x_{n+1},x_n)$
approaches some function $R(x_{n+1},x_n)$ plus a constant proportional
to $N-2$:
\begin{equation} \label {eq:limit-effpair}
\tilde{R}_N(x_{n+1},x_n)\rightarrow R(x_{n+1},x_n)+(N-2)\eta.
\end{equation}
In that case equation (\ref{B7}) follows.
However, it is not clear that (\ref{eq:limit-effpair}) will always be
satisfied. But by imposing a special boundary condition, namely
\begin{equation}
\tilde{R}_2(x_{n+1},x_n)=R(x_{n+1},x_n),
\end{equation}
(\ref{eq:limit-effpair}) will be satisfied exactly.\cite{Chou-Griff} 
The previous boundary condition is the same as saying that the left-most
couple experiences the effective two-particle-potential instead of the true
two-particle-potential. The minimum energy of this system as a function of
the
positions of the two right-most atoms is given by \mbox{$R+N\eta$}.
Assuming
that $R$ is a bounded function, the energy per particle of such a
system will tend to $\eta$ as \mbox{$N\rightarrow \infty$}. $\eta$ is thus
the average energy per particle in any ground state, since the extra
boundary condition only changes the total energy by a term of order 1.
So $R$ is the effective two-particle-potential for the right-most couple of
a
semi-infinite chain. The same is true for $S$ for the left-most couple 
of a semi-infinite
chain extending to the right. $F$ is the total effective
two-particle-potential 
for a couple in a double-infinite chain. $R$,$S$ and $F$ can of course
only be defined up to an additive constant.

In the above derivations the problems arising from the summation of an
infinite number of terms in (\ref{eq:def_right_eff_ener}) are neglected.
In fact, one considers local deformations of length $M$, with the
limit $M\rightarrow \infty$. This will be explained below. Define the
effective two-particle-potential due to the local deformation of length $M$
as 
\begin{eqnarray}
\label{B15}
R^{(M)}(x_{n+1},x_n)\equiv \min_{n+1-M<i<n} \Bigg\{&&\sum_{n+3-M<i\leq
n+1} \left[ K(x_i, x_{i-1}, x_{i-2}) - \eta \right]
\nonumber\\
&&+ \big[ K(x_{n+3-M}, x_{n+2-M}, u_{n+1-M}) \nonumber \\
&&+ K(x_{n+2-M}, u_{n+1-M}, u_{n-M}) - 2 \eta \big] \Bigg\},
\end{eqnarray}
where $u_i$ refers to the ground state value for atom $i$, and where we
have introduced
\begin{equation}
K(x_{n+1},x_n,x_{n-1})\equiv V(x_{n+1})+W(x_{n+1},x_n)+D(x_{n+1},x_{n-1}).
\end{equation}
The right hand site of equation (\ref{B15}) can be rewritten as
\begin{equation}
R^{(M)}(x_{n+1},x_n)=\min_{x_{n-1}} \left[ R^{(M-1)}(x_n,x_{n-1})
+ K(x_{n+1}, x_n, x_{n-1}) - \eta \right].
\end{equation}
It is reasonable to assume that in the limit $M\rightarrow
\infty : R^{(M)}(x_{n+1},x_n)\rightarrow R^{(M-1)}(x_{n+1},x_n)$ (because
$x_{n+2-M}\rightarrow u_{n+2-M}$). Writing $R(x_{n+1},x_n)=
\lim_{M\rightarrow
\infty}R^{(M)}(x_{n+1},x_n)$ the minimization eigenvalue equation
results. 
In the second version of obtaining the equations it is clear that it
is in fact the limit of local deformations, however with the boundary
condition that the left most couple of atoms experiences the effective
two-particle-potential. Here, one should take the length of the chain going
to
infinity in order to let $\eta$ go to the ground state energy per
particle. The above explanation also holds for $S$. It is best to picture
the situation as a local deformation of the ground state.

Now, the nonlinear minimization eigenvalue equations for $R$ and $S$
are rewritten. \cite{Chou-Griff,Griff-stat} 
The eigenvalue equation for $R$ now becomes
\begin{equation} \label{eq:min-eig-R-K}
\eta +R(x_{n+1},x_n)=\min_{x_{n-1}} \left[ R(x_n,x_{n-1})+
K(x_{n+1},x_n,x_{n-1}) \right].
\end{equation}
Let the function $L$ be defined by
\begin{equation}
L(x_{n+1},x_n)=S(x_{n+1},x_n)-V(x_{n+1})-V(x_n)-W(x_{n+1},x_n).
\end{equation}
The minimization eigenvalue equation for $S$ can now be rewritten as
\begin{equation} \label{eq:min-eig-L-K}
\eta
+L(x_{n+1},x_n)=\min_{x_{n+2}} \left[ L(x_{n+2},x_{n+1})
+K(x_{n+2},x_{n+1},x_n) \right].
\end{equation}
In terms of $R$ and $L$ one has
\begin{equation}
F(x_{n+1},x_n)=R(x_{n+1},x_n)+L(x_{n+1},x_n).
\end{equation}

In fact there may be multiple solutions of the eigenvalue equation,
not only differing by a trivial
constant.\cite{Chou-Griff,Griff-stat} The existence of different 
solutions is related to the existence of different degenerate ground
states. The general solution is given by
\begin{equation}
R(x_{n+1},x_n)=\min_{\alpha} \left[ R_{\alpha}(x_{n+1},x_n)+K_{\alpha}
\right].
\end{equation}
The $R_{\alpha}$ correspond to the pure phases and the $K_{\alpha}$
are arbitrary constants.

For each solution of the minimization eigenvalue equation for $R$
(\ref{eq:min-eig-R-K}),
a $\tau$ map can be defined, \cite{Chou-Griff,Griff-stat} where
$\tau(x_{n+1},x_n)=\{(x_n,x_{n-1})\}$ with $x_{n-1}$ one of the values
for which the minimum on the right hand side of
(\ref{eq:min-eig-R-K}) is achieved. An $R$-orbit is defined as
\begin{equation}
\forall n:~ (x_n,x_{n-1})\in \tau(x_{n+1},x_n) ~\Rightarrow~ \eta
+R(x_{n+1},x_n)=R(x_n,x_{n-1})+K(x_{n+1},x_n,x_{n-1}).
\end{equation}
Similarly, for the minimization eigenvalue equation for $L$
(\ref{eq:min-eig-L-K}), a $\sigma$ map can be defined, where
$\sigma(x_{n+1},x_n)=\{(x_{n+2},x_{n+1})\}$ with $x_{n+2}$ one of the
values for which the minimum on the right hand side of
(\ref{eq:min-eig-L-K}) is achieved. An $L$-orbit is defined as
\begin{equation}
\forall n:~ (x_{n+2},x_{n+1})\in \sigma(x_{n+1},x_n) ~\Rightarrow~ \eta
+L(x_{n+1},x_n)=L(x_{n+2},x_{n+1})+K(x_{n+2},x_{n+1},x_n).
\end{equation}
A ground state is both an $R$-orbit and
an $L$-orbit. Therefore, it can be proven that for a ground state
\begin{equation}
F(x_{n+1},x_n) = R(x_{n+1},x_n)+L(x_{n+1},x_n) = F(x_n,x_{n-1})
\end{equation}
So, $F$ is constant on the positions of two adjacent atoms in a ground
state, which is logical since it is the effective two-particle-potential.

Numerical procedures are based on a discretized version of the system. In
that case for each ground
state there is a solution for the eigenvalue equation for which there
is a path from each point to the ground state in the corresponding
$\tau$ graph.\cite{Floria-Griff} So, the situation is as follows. There is
a local deformation of length $M$ (with $M\rightarrow \infty$), in a chain
coinciding with a particular ground state for $\pm \infty$. This ground 
state corresponds
to a certain solution of the eigenvalue equation. The deformation is such
that
atoms $n$ and $n+1$ have values $x_{n}$ and $x_{n+1}$. By applying the
corresponding $\tau$ map one can obtain the positions of the atoms
left from $n$. For the discretized system one will finally reach the
ground state in this way (In the continuous case it is supposed to converge     to the ground state). The same can be done for the atoms right
from $n+1$ by applying the $\sigma$ map. 
From this picture it is clear that the ground state is both an $R$- orbit 
and an $L$-orbit.
In fact the $\tau$ map and
the $\sigma$ map may be multi-valued. So, the atomic positions of
the atoms (say) left from $n$ do not have to be unique. The
deformation can have parts consisting of minimizing cycles (cycles of
minimal energy) different from the ground state configuration at $\pm
\infty$. In the $\tau$ graph one can go directly to the
minimizing cycle corresponding to the ground state configuration at
$-\infty$, or one can first stay for some time in another minimizing
cycle if this exists. If there are
several solutions for $R$ and $S$ (with several corresponding $\tau$ and
$\sigma$ maps), there are several possibilities
to construct $F$. It will often be logical to take the ground states, toward
which the chain converges at $-\infty$ and $+\infty$, the same. The
numerical algorith we used is an adapted version of the one discussed by
Floria and Griffiths. \cite{Floria-Griff}

Here an example will be given to show that it is important that in fact 
limits of
finite deformations are considered. Suppose that the ground state is
ferroelectric, with two degenerate ground states: $x_n= x = \pm l$ where 
$l \ne 0$.
If the deformation is just infinite as suggested in
(\ref{eq:def_right_eff_ener}) the value of $R(l,l)$ should be the same
as the value for $R(-l,-l)$. However, something else is seen. 
Two solutions can be found corresponding to the two ground states. In the
solution corresponding to the solution $x_n=l$, $R(-l,-l)$ has a
higher value than $R(l,l)$. The difference is the defect energy, the
energy cost for going from the $+l$ phase to the $-l$ phase. (The
defect energy (and the defect configuration) can also be calculated using
the $\tau$ map.) From this
it can be seen that the deformation is in fact embedded in the ground
state $u_i=+l$ at $-\infty$ (In the limit $M\rightarrow \infty$:
$x_{n+2-M}\rightarrow u_{n+2-M}$ where $u_{n+2-M}=+l$, or for the second 
version: the
left most couple of atoms in the finite chain experiences the effective
two-particle-potential corresponding to the ground state $u_i=+l$).

It can also be expected that $F$ has local minima at the positions of
two adjacent atoms in metastable states. However, since only two atoms
are at a fixed position, while the other atoms are free to rearrange
themselves in an optimal way, this may not be the case. When
changing the positions of two adjacent atoms by an infinitesimal
amount, the changes of the other atomic positions in the metastable
state does not have to be infinitesimal. Therefore, it is not necessarily
true that there is a local minimum in $F$ for positions of two
adjacent atoms in a metastable state. When there are no other atomic
positions in the metastable state (a period 1 solution), $F$ does have
a local minimum. When the lowest metastable state has positions of two
adjacent atoms which are not seen in a ground state (which will often
be the case), there will be a local minimum in $F$ for these two
positions. In that case the energy cannot be lowered by changing the other
atoms
 by
any amount, since the only states which have lower energy are ground
states and these cannot be reached since the positions of the two
adjacent atoms in consideration are not in a ground state (and the
changes of them should be infinitesimal). By following the development
of the shape of $F$ one may also investigate the kind of phase
transitions that are involved. For example, a discontinuous change in the
set of
points where $F$ achieves its global minimum, indicates a first order
transition.\cite{Griff-stat}

\begin{figure}
\begin{center}
\epsfig{file=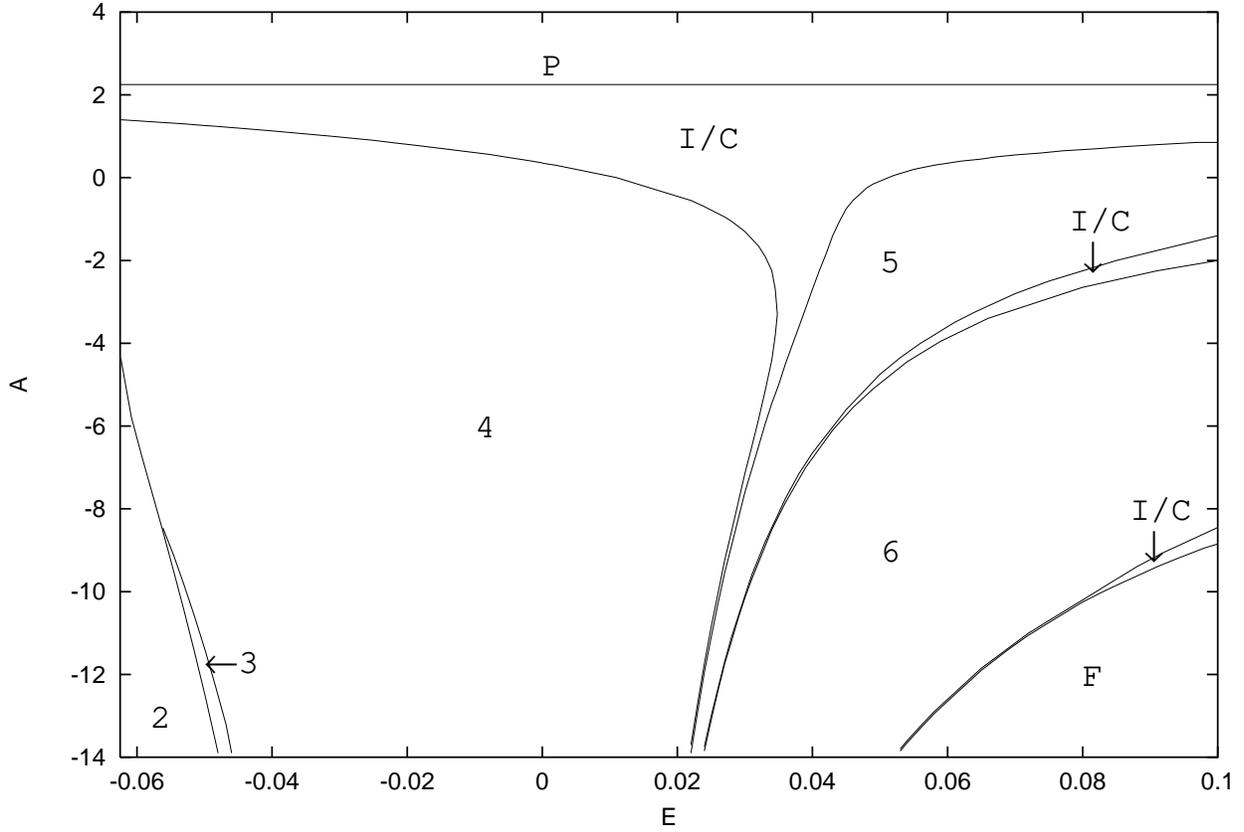,width=\linewidth}
\end{center}
\caption{The phase diagram for the extended DIFFOUR model with $B=1, C=1,
 D=-1$. Shown are the paraelectric phase (P), ferroelectric phase (F),
antiferroelectric phase (2), and commensurate phases with period 3,4,5,6.
Incommensurate phases are labeled I, higher order commensurate
phases C. There is no stable ground state for $E < -1/16$.}
\label{phasediagea}
\end{figure}

\begin{figure}
\begin{center}
\epsfig{file=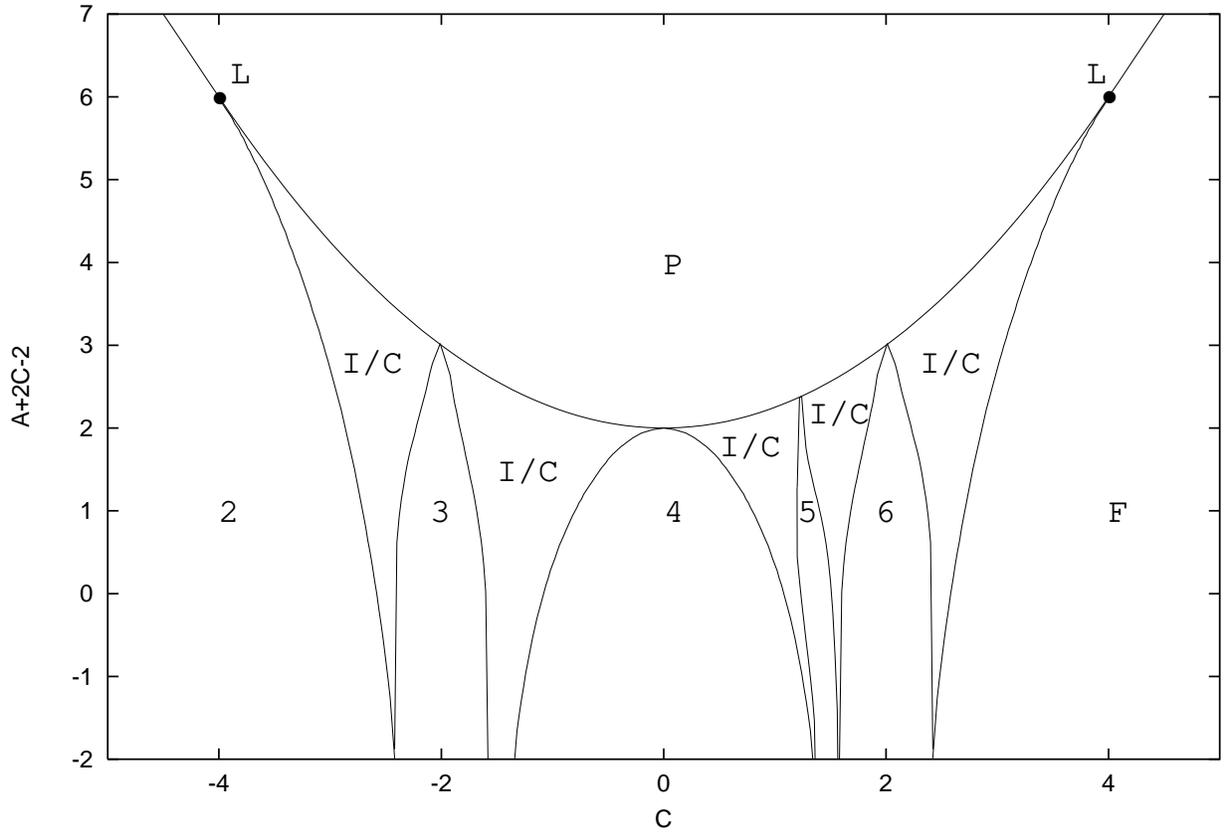,width=\linewidth}
\end{center}
\caption{The phase diagram for the extended DIFFOUR model with $B=1, D=-1$
and $E=0$. This is the original DIFFOUR model. Note the symmetry $C
\leftrightarrow -C$. Same labeling as in figure \ref{phasediagea}. L denotes 
the Lifshitz points.}
\label{phasediagac0}
\end{figure}

\begin{figure}
\begin{center}
\epsfig{file=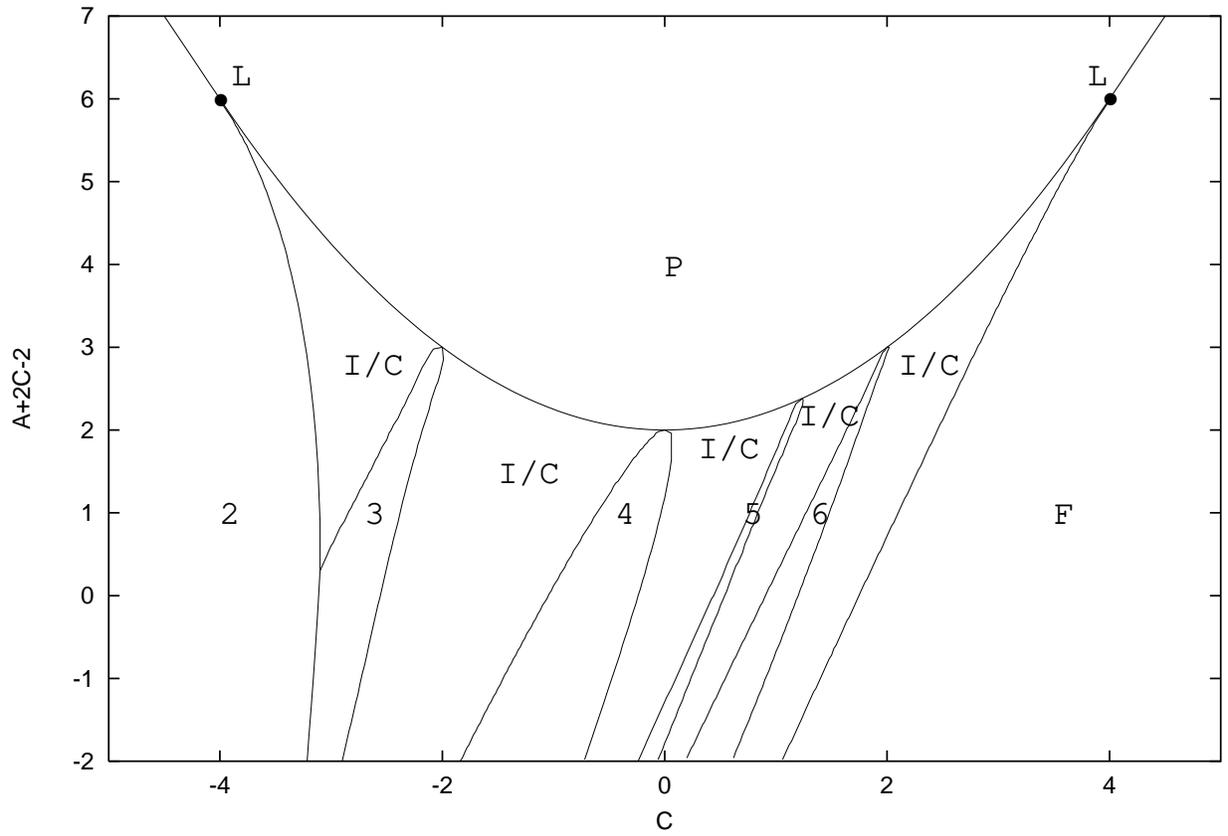,width=\linewidth}
\end{center}
\caption{The phase diagram for the extended DIFFOUR model with $B=1, D=-1$
and $E=1$. Note the asymmetric character, although the boundary of the
paraelectric phase is the same as in figure \ref{phasediagac0}. Same
labeling as in figure \ref{phasediagac0}.} 
\label{phasediagac1}
\end{figure}

\begin{table}
\caption{Stability limits for the paraelectric phase (\ref{parphonon}).
Note that $B + 16 E > 0$ must be satisfied.}
\label{partab}
\begin{tabular}{ccc}
Parameter range & $q$-value of instability & Conditions for having a stable
state \\
\tableline
$D<0$: & & \\
$C < 4 D$ &  $q_c = \pi$ & $A > -4C$ \\
$4D \leq C \leq -4D$ &  $\cos(q_c) = \frac{-C}{4D}$ &  $A + \frac{(4 C + 16
D)^2}{64 D} > 0$ \\
$C > -4D$  & $q_c = 0$ &  $A > 0$ \\
\tableline
$D>0$: & & \\
$C < 0$  & $q_c = \pi$  & $A > -4C$ \\
$C > 0$  & $q_c = 0$  & $A > 0$ \\
\end{tabular}
\end{table}

\begin{table}
\caption{Stability limits for the ferroelectric phase (\ref{ferphonon}).
Note that $B + 16 E > 0$ and $A<0$ must be satisfied.}
\label{fertab}
\begin{tabular}{ccc}
Parameter range & $q$-value of instability & Conditions for having a stable
state \\
\tableline
$D<0$: & & \\
$C < \frac{2AE}{B} + 4D$ & $q_c = \pi$ & $A < 2C - \frac{4AE}{B}$ \\
$\frac{2AE}{B} + 4D \leq C \leq \frac{2AE}{B} - 4D$ & $\cos(q_c) =
\frac{-C}{4D} + \frac{AE}{2BD}$ & $-2A + \frac{(C + 4D - 2AE/B)^2}{4 D} >
0$ \\
$C > \frac{2AE}{B} - 4D$ & $q_c = 0$ & $A < 0$ \\
\tableline
$D>0$: & & \\
$C < \frac{2AE}{B}$ & $q_c = \pi$ & $A < 2C - \frac{4AE}{B}$ \\
$C > \frac{2AE}{B}$ & $q_c = 0$ & $A < 0$ \\
\end{tabular}
\end{table}

\begin{table}
\caption{Stability limits for the antiferroelectric phase
(\ref{antiferphonon}). Both $B + 16 E > 0$ and $A + 4 C < 0$ must be
satisfied.}
\label{antifertab}
\begin{tabular}{ccc}
Parameter range & $q$-value of instability & Conditions for having a stable
state \\
\tableline
$D<0$: & & \\
$C - 10E \frac{A + 4C}{B+16E} < 4D$  & $q_c = 0$ & $A < -4C$ \\
$4D \leq C - 10E \frac{A + 4C}{B+16E} \leq 0$ & $\cos(\frac{q_c}{2}) = 
\frac{C}{4D} - 10E \frac{A + 4C}{4D(B+16E)}$ & $A>-2C-4D+(3B+28E)
\frac{A+4C}{B+16E}$ \\
 & & $-\frac{1}{4D} \left(-C+10E\frac{A+4C}{B+16E} \right)^2$ \\
$0 \leq C - 10E \frac{A + 4C}{B+16E} \leq - 4D$ & $\cos(\frac{q_c}{2}) = 
\frac{-C}{4D} + 10E \frac{A + 4C}{4D(B+16E)}$ & $A>-2C-4D+(3B+28E)
\frac{A+4C}{B+16E}$ \\
 & & $-\frac{1}{4D} \left(-C+10E\frac{A+4C}{B+16E} \right)^2$ \\
$C - 10E \frac{A + 4C}{B+16E} > 4D$  & $q_c = 0$ & $\frac{A}{A+4C} < \frac{3B +
8E}{B + 16E}$ \\
\tableline
$D>0$: & & \\
$C < 10E \frac{A + 4C}{B+16E}$ & $q_c = 0$ & $A < -4C$ \\
$C > 10E \frac{A + 4C}{B+16E}$ & $q_c = 0$ & $\frac{A}{A+4C} < \frac{3B +
8E}{B + 16E}$ \\
\end{tabular}
\end{table}


\end{document}